\documentclass[12pt,a4paper]{article}
\usepackage[cp1251]{inputenc}
\usepackage[english]{babel}
\usepackage{epsfig}
\usepackage{amssymb}
\usepackage{amsmath}

\newcommand{\bm}[1]{{\boldsymbol{\bf #1}}}
\renewcommand{\vec}[1]{{\bm{#1}}}
\newcommand{\fr}[2]{{\dfrac{#1}{#2}}}
\newcommand{\sfr}[2]{{{#1}/{#2}}}

\newcommand{\pdiff}[2]{{\fr{\partial{#1}}{\partial{#2}}}}

\newcommand{\spdiff}[2]{{\sfr{\partial{#1}}{\partial{#2}}}}

\title%
{%
Formation of Accretion Disks in Close-Binary Systems with Magnetic Fields%
\footnote{Submitted in Astronomy Reports, 2010, \textbf{54}, No. 12.}
}%
\author%
{%
 A.G. Zhilkin$^{1,2}$\thanks{E-mail: zhilkin@inasan.ru}, \ %
 D.V. Bisikalo$^{1}$\\ %
\textit{\small $^{1}$ Institute of Astronomy, Russian Academy of Sciences, Moscow, Russia}\\%
\textit{\small $^{2}$ Chelyabinsk State University, Chelyabinsk, Russia}\\%
}%
\date{}

\begin{document}

\maketitle%

\begin{abstract}

\noindent We have developed a three-dimensional numerical model and applied it to simulate plasma flows in semi-detached binary systems whose accretor possesses a strong intrinsic magnetic field. The model is based on the assumption that the plasma dynamics are determined by the slow mean flow, which forms a backdrop for the rapid propagation of MHD waves. The equations describing the slow motion of matter were obtained by averaging over rapidly propagating pulsations. The numerical model includes the diffusion of magnetic field by current dissipation in turbulent vortices, magnetic buoyancy, and wave MHD turbulence. A modified three-dimensional, parallel, numerical code was used to simulate the flow structure in close binary systems with various accretor magnetic fields, from $10^5$ to $10^8$~G. The conditions for the formation of the accretion disk and the criteria distinguishing the two types of flow corresponding to intermediate polars and polars are discussed.
\end{abstract}

\section{Introduction}

Among close-binary systems in which magnetic fields exert a substantial influence on the flow of matter in the system, the most prominent are polars and intermediate polars \cite{Warner1995}. These are semi-detached binary systems consisting of a low-mass, late-type star (the donor) and a white dwarf (the accretor). During the process of mass transfer, matter flows from the donor onto the white dwarf through the inner Lagrange point.

In polars (AM Her-type systems), the white dwarf has a strong intrinsic magnetic field ($B_a \approx 10^7-10^8$~G at the surface). These systems are characterized by relatively short orbital periods (1-5 hr),and the rotation of the components is synchronized \cite{Norton2004}. Analysis of observational data shows that accretion disks do not form in polars; instead, the matter flowing from the donor forms a collimated flow, which falls onto a region around one of the magnetic poles of the accretor along the star's magnetic-field lines \cite{Warner1995, Campbell1997}.

The white dwarfs in intermediate polars possess relatively weak magnetic fields ($B_a \approx 10^4-10^6$~G at the surface). These systems occupy an intermediate position between polars and non-magnetic cataclysmic variables. The rotational periods of the accretors in these systems are appreciably shorter than the orbital periods, by factors of one, two, or even three orders of magnitude \cite{Norton2004}. The asynchronicity of the rotation of the accretor in intermediate polars can  be explained as an effect of the interaction of the magnetic field of the white dwarf with the disk material near the boundary of the magnetosphere. As a result of this interaction, an equilibrium rotational regime is established in which the corotation radius is equal to the magnetosphere radius \cite{Warner1995, Lipunov1987}.

Our previous studies \cite{Zhilkin2009, ZhilkinASR2010, Zhilkin2010} have used three-dimensional numerical simulations to investigate the flow structures in close-binary systems (using SS Cyg as an example), allowing for relatively weak, dipolar accretor magnetic fields ($B_a = 10^5$~G at the surface). Modeling with high values of $B_a$ is more difficult due to the fact that increasing the field strength leads to rigid constraints on the time step required for the difference scheme to be stable (the Courant-Friedrichs-Levy condition). However, a simple analysis shows that this problem has a fundamental, rather than technical, nature. Indeed, the Alfv{\'e}n velocity, estimated by the expression
\begin{equation}\label{eq1.1}
 \fr{u_A}{c} = 2.97 \left( \fr{B_a}{10^7~\text{G}} \right)
 \left( \fr{\rho}{10^{-9}~\text{g/cm}^3} \right)^{-1/2},
\end{equation}
can exceed the speed of light $c$ in the accretion flows of polars. Therefore, in general, it is not correct to
use a non-relativistic magnetic gas-dynamical approximation when modeling the structure of such a flow. However, it does not really make sense to apply relativistic magnetic gas-dynamics in this case, since the plasma flow itself is appreciably non-relativistic. Only Alfv{\'e}n and magneto-acoustic waves in the accretion flow will propagate with relativistic speeds. Over the characteristic dynamical time for the evolution of the slow plasma flow, the MHD waves are able to pass through the accretion flow many times (in the longitudinal and transverse directions). As a result, the dynamics of the plasma flow can be treated in a modified, non-relativistic, magnetic, gas-dynamical framework, as a mean flow that forms the backdrop for wave MHD turbulence.

We propose here a numerical model for computing the flow structures in close-binary systems with strong accretor magnetic fields. We present results of our numerical simulations obtained using a modified, three-dimensional, parallel code. These computations were carried out for magnetic fields at the accretor surface from $10^5$ to $10^8$~G.

The paper is organized as follows. Section 2 describes the main equations and numerical method used in the framework of the modified model. Section 3 presents a comparison of the flow structures in a semi-detached system similar to SS Cyg for various white-dwarf magnetic fields, based on the three-dimensional numerical simulation results. The Conclusion briefly discusses the main results of our study.

\section{Numerical method}

\subsection{Main equations}

We will use a non-inertial reference frame that rotates together with the binary system with the angular velocity $\vec{\Omega}$ relative to its center of mass. We will use the Cartesian coordinates ($x$, $y$, $z$) in this frame, with the coordinate origin coincident with the center of the accretor and the center of the donor located a distance $A$ from the accretor along the $x$ axis. The $z$ axis is directed along the rotational axis of the system.

Let us consider a semi-detached binary system whose accretor rotates synchronously, so that the rotational period of the accretor is equal to the orbital period of the system. We will take the magnetic field of the accretor to be dipolar, so that the vector magnetic field is given by
\begin{equation}\label{eq2.1}
 \vec{B}_{*} = 
 \fr{3(\vec{\mu} \cdot \vec{r}) \vec{r}}{r^5} -
 \fr{\vec{\mu}}{r^3}.
\end{equation}
In general, the direction of the magnetic moment $\vec{\mu}$ may not coincide with the direction of the angular velocity of the binary system $\vec{\Omega}$. In our chosen coordinate system, the vector $\vec{\mu}$ has the components $\mu_x = \mu \sin\theta \cos\phi$, $\mu_y = \mu \sin\theta \sin\phi$, $\mu_z = \mu \cos\theta$, where $\mu$ is the modulus of $\vec{\mu}$, $\theta$ is the inclination of $\vec{\mu}$ to the $z$ axis, and $\phi$ is the angle between the $x$ axis and the projection of $\vec{\mu}$ onto the $xy$ plane. Note that a magnetic field of the form \eqref{eq2.1} is potential $\nabla \times \vec{B}_{*} = 0$ and, in the case of synchronous rotation, stationary $\spdiff{\vec{B}_{*}}{t} = 0$.

The magnetic field of the accretor can be fairly strong in the region of the magnetosphere. Therefore, it is convenient to represent the total magnetic field in the plasma $\vec{B}$ as the sum of the accretor magnetic field $\vec{B}_{*}$ and the magnetic field induced by currents in the plasma itself $\vec{b}$: $\vec{B} = \vec{B}_{*} + \vec{b}$. Using the properties of the field $\vec{B}_{*}$ noted above, it can partially be eliminated from the magnetic gas-dynamical equations \cite{Tanaka1994, Powell1999, Kulikovsky2001}.

We described plasma flows arising due to mass transfer in the close-binary system, taking into account the strong magnetic field of the compact object $\vec{B}_{*}$, using the system of equations
\begin{equation}\label{eq2.2}
 \pdiff{\rho}{t} + \nabla \cdot \left( \rho\vec{v} \right) = 0,
\end{equation}
\begin{equation}\label{eq2.3}
 \pdiff{\vec{v}}{t} + \left(\vec{v} \cdot \nabla \right) \vec{v} =
 -\fr{\nabla P}{\rho} -
 \fr{\vec{b} \times (\nabla \times \vec{b})}{4\pi\rho} + 
 2 (\vec{v} \times \vec{\Omega}) + \vec{g} -
 \fr{\vec{v}_{\perp}}{\tau},
\end{equation}
\begin{equation}\label{eq2.4}
 \pdiff{\vec{b}}{t} =
 \nabla \times \left[
  \vec{v} \times \vec{b} + 
  \vec{v} \times \vec{B}_{*} -
  \eta (\nabla \times \vec{b}) 
 \right],
\end{equation}
\begin{equation}\label{eq2.5}
 \rho T \left[
 \pdiff{s}{t} + (\vec{v} \cdot \nabla) s
 \right] = 
 n^2 \left(\Gamma - \Lambda\right) + 
 \fr{\eta}{4\pi}(\nabla \times \vec{b})^2.
\end{equation}
Here, $\rho$ is the density, $\vec{v}$ the velocity, $P$ the pressure, $s$ the entropy per unit mass of the gas, $n = \rho/m_{\text{p}}$ the number density, $m_{\text{p}}$ the proton mass, $\eta$ the magnetic viscosity, and $\vec{g}=-\nabla\Phi$, with $\Phi$ being the Roche potential. The effects of radiative heating and cooling are included in the entropy equation \eqref{eq2.5}, as well as heating of matter due to current dissipation (the last term). The radiative heating and cooling functions $\Gamma$ and $\Lambda$ have complex dependences on the temperature $T$ \cite{Cox1971, Dalgarno1972, Raymond1976, Spitzer1981}. Our numerical model uses a linear approximation for these functions in the vicinity of the equilibrium temperature, $T = 11230~\text{K}$ \cite{Bisikalo2003, Zhilkin2009, ZhilkinMM2010}, corresponding to an effective temperature of the accretor of 37000~K. The term $2\left(\vec{v} \times \vec{\Omega}\right)$ in the equation of motion \eqref{eq2.3} describes the Coriolis force. The density, entropy, and pressure are related by the equation of state of an ideal gas: $s = c_V \ln(P / \rho^{\gamma})$, where $c_{V}$ is the specific heat of the gas at constant volume and $\gamma = 5/3$ is the adiabatic index.

The numerical model includes the effects of magnetic-field diffusion [see \eqref{eq2.4}]. The earlier analysis of \cite{Zhilkin2009, ZhilkinASR2010} showed that two effects dominate in accretion disks forming in close-binary systems \cite{Campbell1997}. The first is magnetic reconnection and current dissipation in turbulent vortices. The second is buoyancy of force tubes of the toroidal magnetic field generated in the disk due to its differential rotation. Accretion disks may not be able to form in systems with strong magnetic fields, in which case the flow will have the form of an accretion-column stream directed from the inner Lagrange point $L_1$ toward one of the
magnetic poles of the accretor (a polar). In this case, the main effect leading to current dissipation in the plasma is probably wave MHD turbulence due to the propagation of Alfv{\'e}n and magneto-acoustic waves in the accretion-column stream. The velocity of these waves will far exceed the velocity of the plasma itself and, in some cases, may even be relativistic. The wave magnetic viscosity (see Appendix B) $\eta_w$ can be obtained from the expression
\begin{equation}\label{eq2.6}
 \eta_w = \alpha_w \fr{l_w B_{*}}{\sqrt{4\pi\rho}},
\end{equation}
where $l_w$ is the characteristic scale for wave pulsations, which can be estimated as the scale for inhomogeneities of the accretor magnetic field: $l_w = B_{*}/|\nabla B_{*}|$. The parameter $\alpha_w$ determines the efficiency of wave diffusion. The total magnetic viscosity $\eta$ due to all effects depends on the magnetic field in the plasma. Therefore, on the whole, the diffusion of the magnetic field has a non-linear character.

The last term in the equation of motion \eqref{eq2.3} describes the force exerted on the plasma by the accretor magnetic field, which influences the component of the plasma velocity perpendicular to the magnetic field lines $\vec{v}_{\perp}$. The basis for this term is provided in Appendix A. Note that the motion of plasma particles across the magnetic field is due primarily to the gravitational force of the compact object (gravitational drift) (see, for example, \cite{FrankKamenetsky1968, Chen1987, Trubnikov1996}). Due to the Larmor character of the motion of particles in the magnetic field, their mean motion is decelerated in the perpendicular direction. The strong external magnetic field plays the role of an effective fluid with which the plasma interacts. The last term in \eqref{eq2.3} can be interpreted as a frictional force between the plasma and magnetic field, whose form is analogous to the friction between the components in a plasma consisting of several types of particles \cite{FrankKamenetsky1968}. Recall that our model assumes synchronous rotation of the accretor. In this case, the velocity of the magnetic field lines in our chosen system is zero. The characteristic time for decay of the perpendicular velocity is
\begin{equation}\label{eq2.7}
 \tau = \fr{4\pi\rho\eta_w}{B_{*}^2}.
\end{equation}
This quantity is determined by wave dissipation of the magnetic field, which is characterized by the diffusion coefficient $\eta_w$ \eqref{eq2.6}. A similar expression for the electro-magnetic force due to the strong magnetic field of an accretor was used in \cite{King1993, Wynn1995, Wynn1997, King1999, Norton2004, Ikhsanov2004, Norton2008}, where the flow structure was modeled using a quasi-particle method. The quasi-particles were taken to be individual blobs of plasma that fall into the Roche lobe of the accretor through the inner Lagrange point $L_1$. In these studies, an expression for the decelerating force of the form $-\sfr{\vec{v}_{\perp}}{\tau}$ was justified by the fact that diamagnetic effects arise in the plasma blobs during their motion in the external magnetic field \cite{Drell1965}, which hinder their free motion perpendicular to the magnetic field lines.

\subsection{Numerical method}

Suppose we know the distribution of all quantities in the computational region at time $t^{n}$. To obtain their values at the next time step, corresponding to time $t^{n+1} = t^{n} + \Delta t$, we split the equations \eqref{eq2.2}--\eqref{eq2.5} according to the various physical processes they represent. The full algorithm consists of five successive steps, which we describe briefly below.

In the first step, a subsystem of equations describing the dynamics of the plasma in its own magnetic field is distinguished:
\begin{equation}\label{eq2.8}
 \pdiff{\rho}{t} + \nabla \cdot \left( \rho\vec{v} \right) = 0,
\end{equation}
\begin{equation}\label{eq2.9}
 \pdiff{\vec{v}}{t} + \left(\vec{v} \cdot \nabla \right) \vec{v} =
 -\fr{\nabla P}{\rho} -
 \fr{\vec{b} \times (\nabla \times \vec{b} )}{4\pi\rho},
\end{equation}
\begin{equation}\label{eq2.10}
 \pdiff{\vec{b}}{t} = \nabla \times \left( \vec{v} \times \vec{b} \right),
\end{equation}
\begin{equation}\label{eq2.11}
 \pdiff{s}{t} + (\vec{v} \cdot \nabla) s = 0.
\end{equation}
The form of this system coincides with the equations of ideal magnetic gas dynamics \cite{Landau2001}. The system can be solved numerically using the higher-order Gudonov-type difference scheme described in \cite{ZhilkinMM2010} (if $\vec{B}_{*} = 0$ in all expressions there). When constructing the difference scheme, we applied the technique of unified variables for MHD equations \cite{Zhilkin2007}, which enabled us to use an adaptive mesh in the numerical code. The computations presented below used a geometrically adaptive mesh that became more dense toward the equatorial plane and the surface of the accretor. This made it possible to appreciably enhance the spatial resolution of the vertical structure of the accretion disk and in the region of the accretor magnetosphere. We used the eight-wave method to clean the divergence of the plasma magnetic field $\vec{b}$ \cite{Powell1999, Dellar2001}. Note that the external magnetic field $\vec{B}_{*}$ does not appear in this system of equations. Therefore, the limitations on the time step $\Delta t$ imposed by the stability condition for the difference scheme (the Courant-Friedrichs-Levy condition) will not be excessively severe.

In the second step, we take into account variations of the gas velocity due to external forces (the Coriolis force and gradient of the Roche potential):
\begin{equation}\label{eq2.12}
 \pdiff{\vec{v}}{t} = 2 (\vec{v} \times \vec{\Omega}) + \vec{g}.
\end{equation}
All remaining quantities are taken to be constant in this step of the algorithm. Since the Roche potential is time-independent, the solution of this equation in the interval $t^n \le t \le t^{n+1}$ can be written
\begin{eqnarray}
 v_x & = & A \cos(2\Omega t) + B \sin(2\Omega t) + \fr{g_y}{2\Omega}, \label{eq2.13a} \\
 v_y & = & B \cos(2\Omega t) - A \sin(2\Omega t) - \fr{g_x}{2\Omega}, \label{eq2.13b} \\
 v_z & = & v_z^0 - g_z (t - t^n). \label{eq2.13c}
\end{eqnarray}
The integration constants in these equations $A$ and $B$ are determined from the initial conditions:
\begin{eqnarray}
 A & = & 
 \left( v_x^0 - \fr{g_y}{2\Omega} \right) \cos(2\Omega t^n) - 
 \left( v_y^0 + \fr{g_x}{2\Omega} \right) \sin(2\Omega t^n),
 \label{eq2.14a} \\
 B & = & 
 \left( v_x^0 - \fr{g_y}{2\Omega} \right) \sin(2\Omega t^n) + 
 \left( v_y^0 + \fr{g_x}{2\Omega} \right) \cos(2\Omega t^n).
 \label{eq2.14b}
\end{eqnarray}
The initial velocities $\vec{v}^0$ are equal to the values obtained at time $t^{n+1}$ in the previous step of the algorithm from the solution of \eqref{eq2.8}--\eqref{eq2.11}.

The third step of the algorithm takes into account deceleration during motion of the plasma across magnetic field lines, as well as the generation of magnetic field due to this motion. The corresponding equations can be written
\begin{equation}\label{eq2.15}
 \pdiff{\vec{v}_{\perp}}{t} = -\fr{\vec{v}_{\perp}}{\tau}, \ \ 
 \pdiff{\vec{b}}{t} = \nabla \times \left( \vec{v}_{\perp} \times \vec{B}_{*} \right).
\end{equation}
Integration of these equations yields
\begin{equation}\label{eq2.16}
 \vec{v}_{\perp} = \vec{v}_{\perp}^0 e^{-\frac{t-t^n}{\tau}}, \ \ 
 \vec{b} = \vec{b}^0  + \tau \left( 1 - e^{-\frac{t-t^n}{\tau}} \right)
 \nabla \times \left( \vec{v}_{\perp}^0 \times \vec{B}_{*} \right).
\end{equation}

In the fourth step, we take into account the effects of magnetic-field diffusion. Recall that the equation describing diffusion of the magnetic field is non-linear in our model. Therefore, the application of explicit methods to solve this equation would lead to excessively severe limitations on the time step. In our approach, this equation was solved numerically using an implicit, locally one-dimensional method with a factorizable operator \cite{Samarsky1971}. We applied regularization of the factorizable operator to deal with the mixed derivatives that arise due to the use of an adaptive mesh. The regularization procedure essentially reduces to replacing the multiplicative operators making up the original factorizable operator with equivalent tridiagonal operators. The regularization parameter is determined by the maximum-modulus eigenvalue of the metric tensor describing the curvilinear coordinate system. To correctly take into account non-linear terms in the scheme, we devised an iterative process that was applied until we obtained a solution with a specified accuracy. In each iteration, a system of linear, algebraic equations with a tridiagonal matrix arises, which is solved numerically using a scalar fitting method. The method used to solve the magnetic-field diffusion equation is described in more detail in \cite{ZhilkinZST2010}.

Finally, the fifth step takes into account the effects of radiative heating and cooling, as well as heating due to current dissipation. These processes are described by the right-hand side of \eqref{eq2.5}. We emphasize that we used semi-analytical or implicit, absolutely stable methods in all other steps of the algorithm (apart from the first). Therefore, no constraints are imposed on the time step $\Delta t$ apart from the Courant-Friedrichs-Levy stability condition [which arises during the solution of \eqref{eq2.8}--\eqref{eq2.11}].

\section{Computational results}

\subsection{Parameters of the model}

Let us consider the formation of an accretion disk in a close-binary system whose parameters correspond to SS Cyg as a function of the accretor magnetic field. The donor of the SS Cyg system is a red dwarf with mass $0.56~M_{\odot}$ and the accretor is a white dwarf with mass $M_a = 0.97~M_{\odot}$. The orbital period of the system is $P_{\text{orb}} = 6.6$~hr, and its semi-major axis is $A = 2.05R_{\odot}$ \cite{Giovannelli1983}. In our computations, the magnetic field at the white-dwarf surface was varied from $10^5$ to $10^8$ G, and the orientation of the magnetic axis was fixed, $\theta=30^{\circ}$ and $\phi=0^{\circ}$.

At the inner Lagrange point $L_1$, we specified the gas velocity to be equal to the local sound speed, $c_s = 7.4~\text{km}/\text{s}$, which corresponds to a donor temperature of 4000~K. The gas density at $L_1$ is $\rho(L1) = {1.1\times 10^{-7}}~\text{g}/\text{cm}^{3}$, and the mass transfer rate is $\dot{M} = 10^{-9}~M_{\odot}/\text{yr}$. We specified the following boundary conditions at the remaining boundaries of the computational domain: density $\rho_{\text{b}} = 10^{-6} \rho(L_1)$, temperature $T_{\text{b}} = 11230~K$, velocity $\vec{v}_{\text{b}} = 0$, and magnetic field $\vec{B}_{\text{b}} = \vec{B}_{*}$. The accretor was taken to be a sphere with radius $0.0125 A$, at whose boundary we specified a free-outflow condition. The velocity of the inflowing plasma was taken to be parallel to $\vec{B}_{*}$. All the matter in a cell occupied by the accretor was taken to fall onto the white dwarf. We emphasize that these boundary conditions were used at each individual stage in the subdivided numerical algorithm described in the previous section. The initial conditions in the computational region were density $\rho_0 = 10^{-6} \rho(L_1)$, velocity $\vec{v}_0 = 0$, temperature $T_0 = 11230~K$, and magnetic field $\vec{B}_0 = \vec{B}_{*}$. We found our solution in the region ($-0.56A \le x \le 0.56A$, $-0.56A \le y \le 0.56A$, $-0.28A \le z \le 0.28A$) on the geometrically adaptive mesh \cite{ZhilkinMM2010}.

\begin{figure}[t]
\centering
\includegraphics[width=0.75\textwidth]{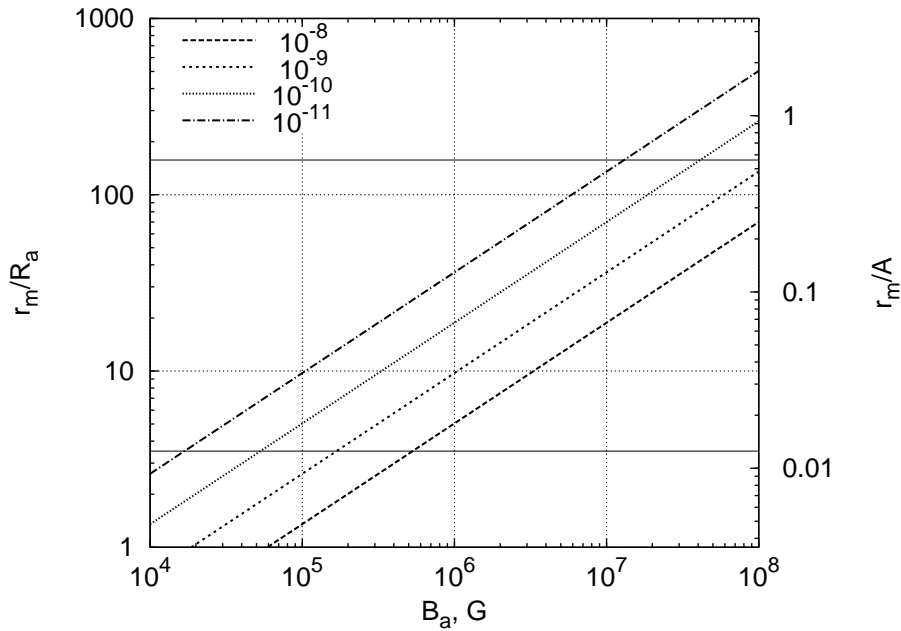}%
\caption{Radius of the white-dwarf magnetosphere in the SS Cyg system as a function of the surface magnetic field. The individual curves correspond to various accretion rates (in units of $M_{\odot}/\text{yr}$). The solid horizontal lines show the numerical boundary of the accretor surface, $0.0125A$ (lower), and the distance from the accretor center to the inner Lagrange point, $0.56A$ (upper).}%
\label{fg1}
\end{figure}

The extent to which the magnetic field influences the flow structure can be determined using estimates of the radius of the accretor magnetosphere. We will assume that the magnetic pressure is equal to the dynamical pressure of the accreting gas at the boundary of the magnetosphere (see \cite{Lipunov1987}):
\begin{equation}\label{eq3.1}
 \fr{B^2}{8\pi} = \rho v_{ff}^2,
\end{equation}
where $v_{ff} = \sqrt{\sfr{2GM_a}{r}}$ is the free-fall velocity. We can find the density of the matter from the expression for the accretion rate: $\dot{M}_a = 4\pi r^2 \rho v_{ff}$. Substituting $B = B_a (\sfr{R_a}{r})^3$, where $R_a$ is the radius of the accretor, into \eqref{eq3.1}, we can find the radius of the magnetosphere:
\begin{equation}\label{eq3.2}
 r_m = \left( \fr{B_a^4 R_a^{12}}{8 G M_a \dot{M}_a^2} \right)^{\frac{1}{7}}.
\end{equation}
The dependence of the magnetosphere radius $r_m$ on the magnetic field $B_a$ at the accretor surface is shown in Fig. \ref{fg1}. The different lines correspond to different accretion rates $\dot{M}_a$ in units of $M_{\odot}/\text{yr}$. The solid horizontal lines show the specified accretor radius, $0.0125A$ (lower line), and the distance from the center of the accretor to the inner Lagrange points, $0.56A$ (upper line). In our computations, the characteristic accretion rate was of the order of $10^{-10} M_{\odot}/\text{yr}$. In this case, the magnetosphere radius is approximately $5-6 R_a$ for $B_a = 10^5$~G and roughly $20 R_a$ for $B_a = 10^6$~G. When $B_a = 10^4$~G, the magnetosphere radius becomes smaller than the radius of the ''numerical star'', although it remains larger than the radius of the white dwarf.

We present results for eight computational models here, with magnetic fields at the accretor surface equal to $10^5$~G (model 1), ${5 \times 10^5}$~G (model 2), $10^6$~G (model 3), ${5 \times 10^6}$~G (model 4), $10^7$~G (model 5), ${5 \times 10^7}$~G (model 6), and $10^8$~G (model 7). These models can be divided into two groups. The first includes models 1, 2, and 3, with relatively weak magnetic fields, in which accretion disks form. These models presumably correspond to the case of intermediate polars. The second group contains models 4, 5, 6, and 7, with strong magnetic fields and without the formation of an accretion disk. These models correspond to the case of polars. The computations for all the models were continued until a quasi-stationary regime was reached, when the total mass of matter in the computational domain was roughly constant (to within 1\%). In models in which an accretion disk formed, the time for establishing this quasi-stationary regime was about 10-15 orbital periods. In models with strong magnetic fields and without the formation of an accretion disk, the transition to a quasi-stationary regime was more rapid (about five orbital periods). We will describe the results for these two groups of models separately.

\subsection{Weak magnetic field}

\begin{figure}[ht]
\centering
\includegraphics[height=7cm]{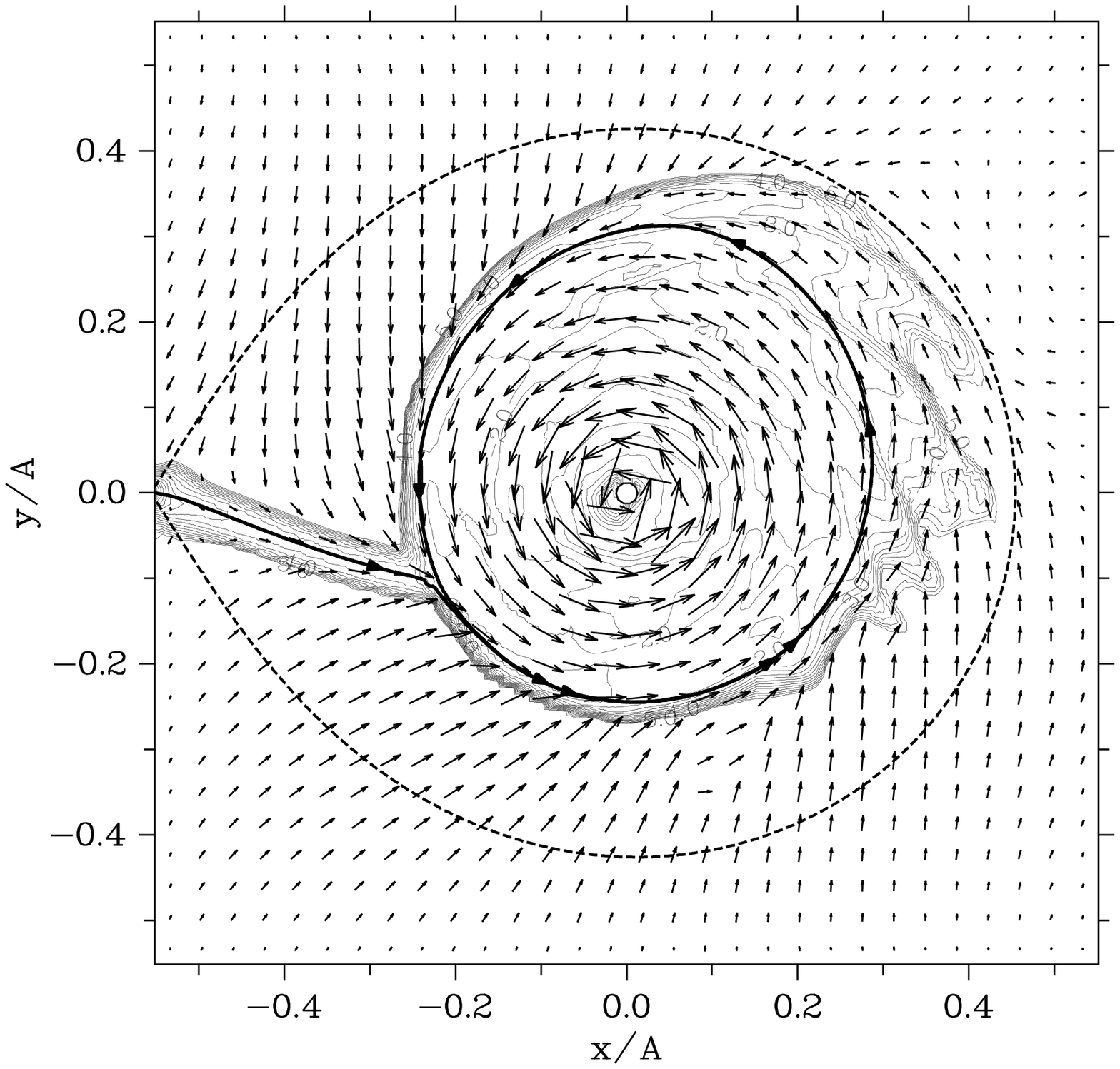}%
\hspace{0.25cm}
\includegraphics[height=7cm]{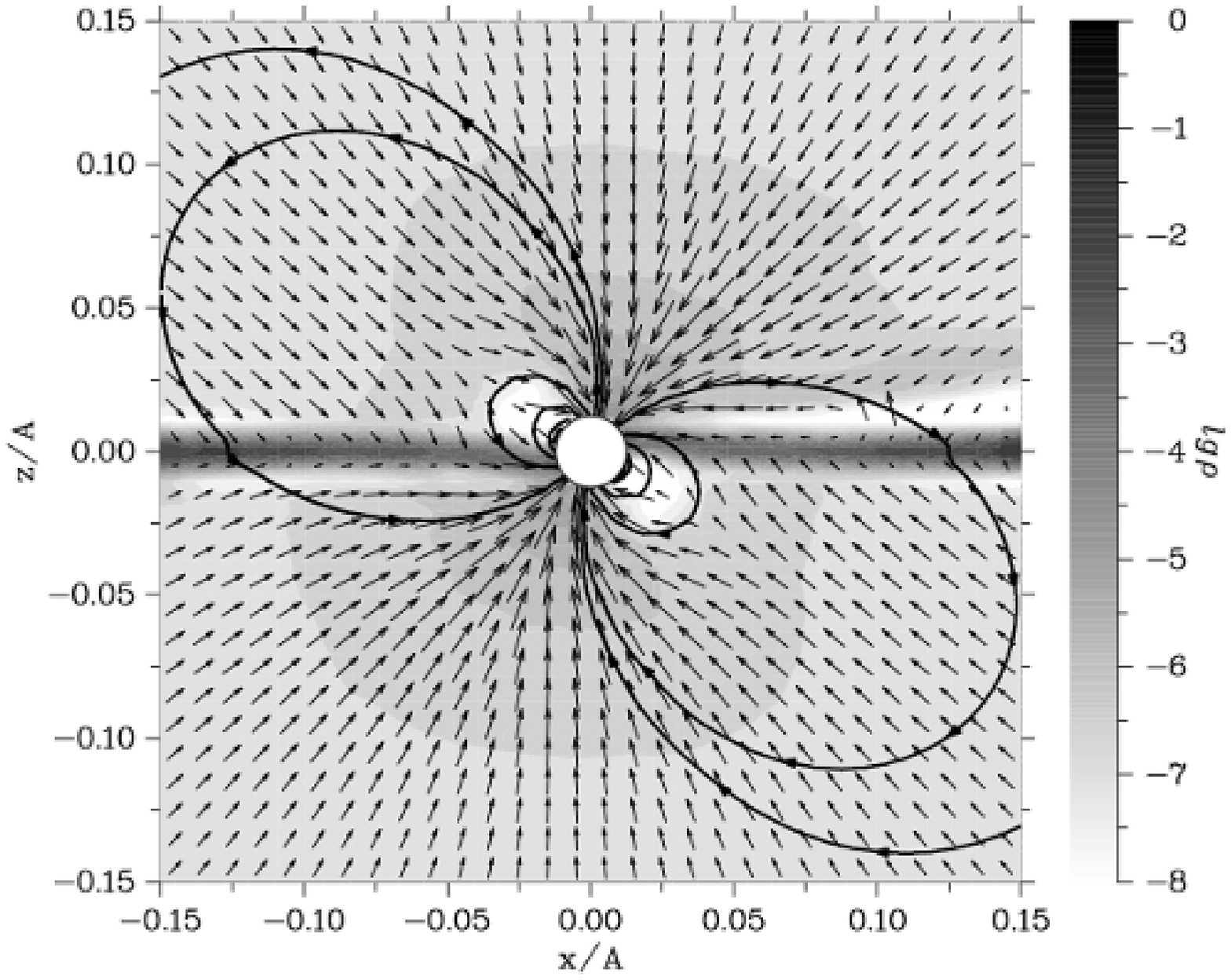}%
\caption{Flow structure in the equatorial (left) and vertical (right) planes for model 1 ($B_a = 10^5$~G). The distributions of the density (contours on the left and gray scale on the right) and velocity (arrows) are shown. The dashed curve corresponds to the boundary of the accretor Roche lobe. The bold curves with arrows show either the streamlines from $L_1$ (left) or the magnetic lines (right).}%
\label{fg2}
\end{figure}

\begin{figure}[ht]
\centering
\includegraphics[height=7cm]{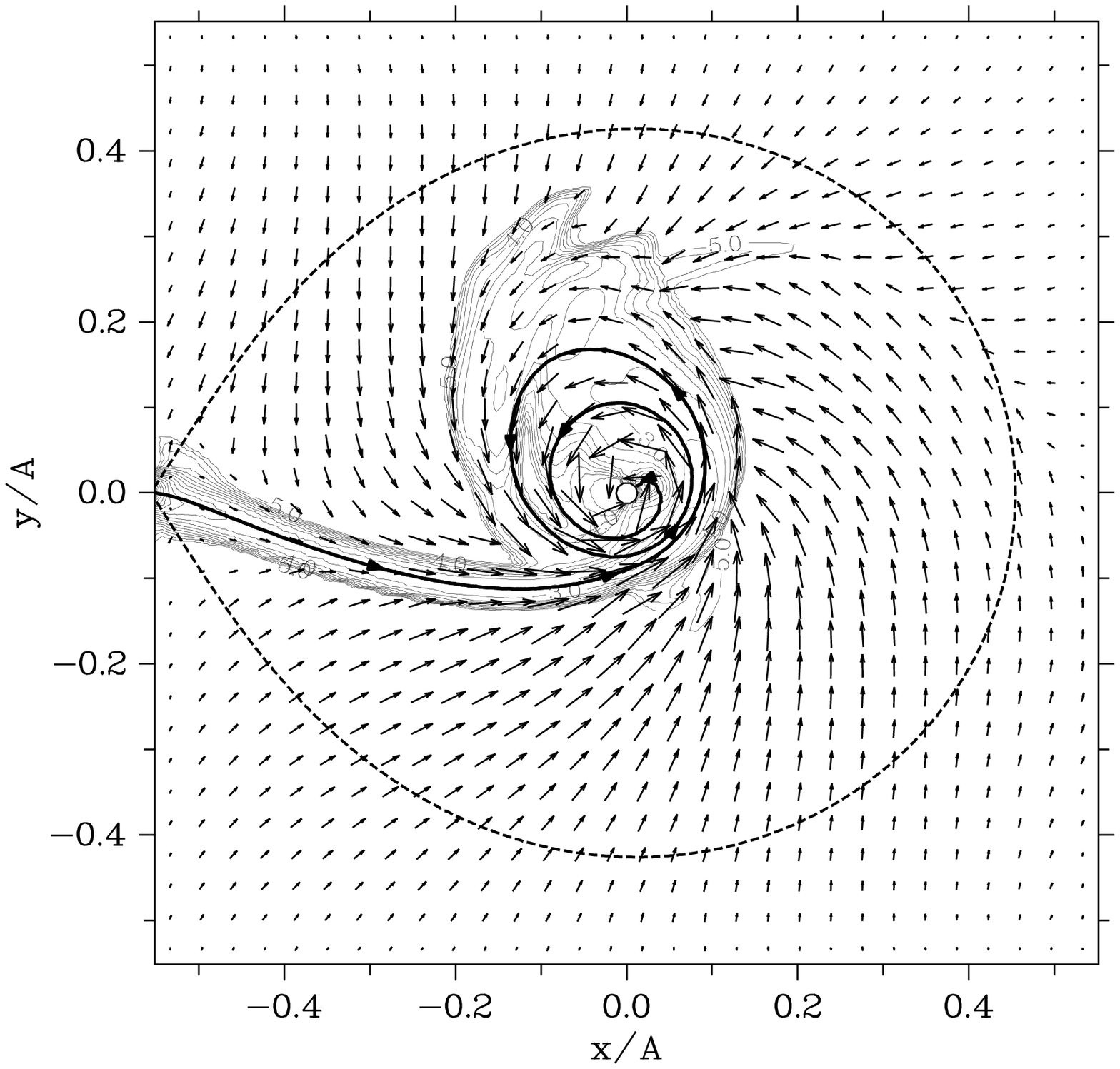}%
\hspace{0.25cm}
\includegraphics[height=7cm]{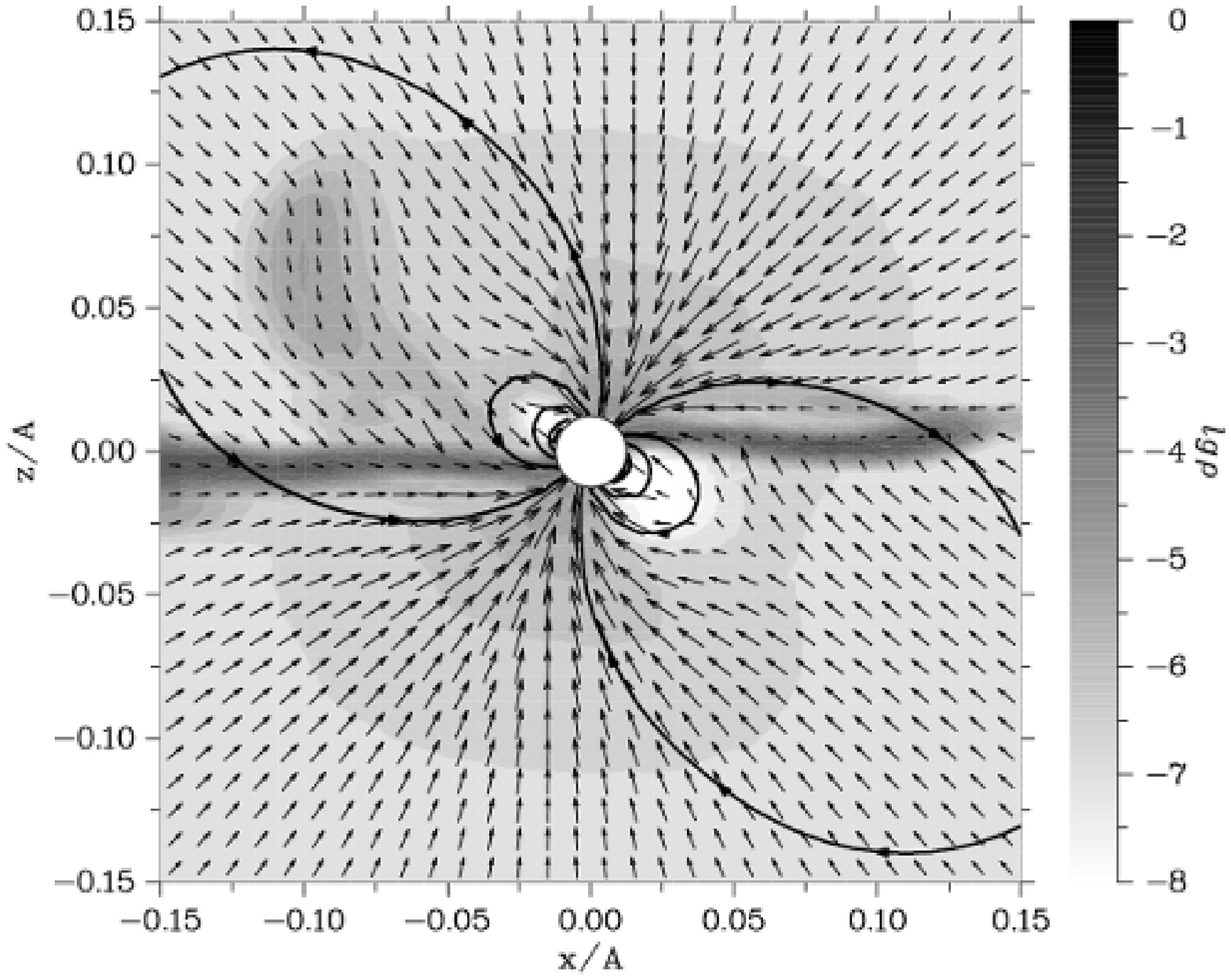}%
\caption{Same as Fig. \ref{fg2} for model 2 ($B_a = {5 \times 10^5}$~G).}%
\label{fg3}
\end{figure}

\begin{figure}[ht]
\centering
\includegraphics[height=7cm]{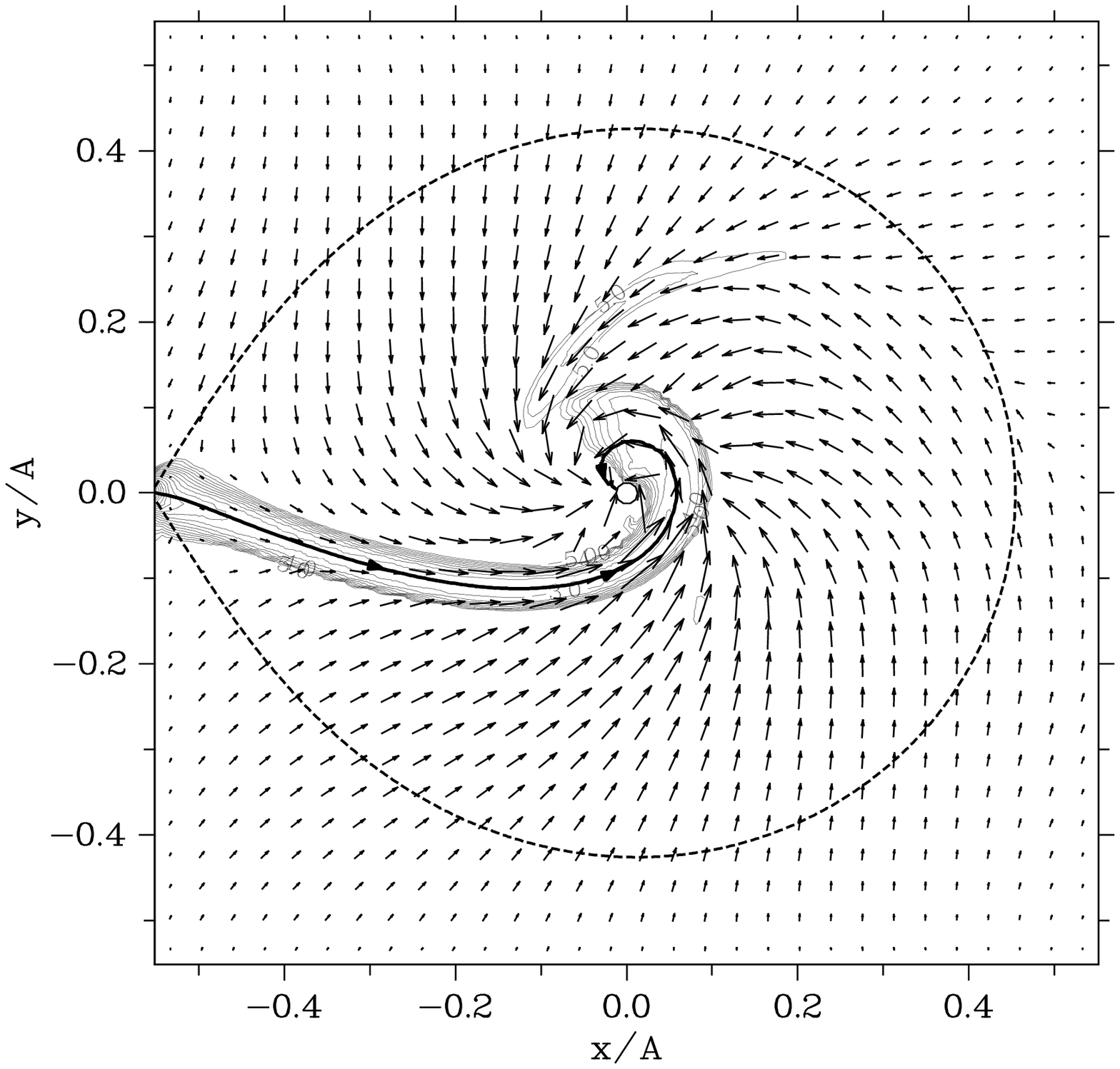}%
\hspace{0.25cm}
\includegraphics[height=7cm]{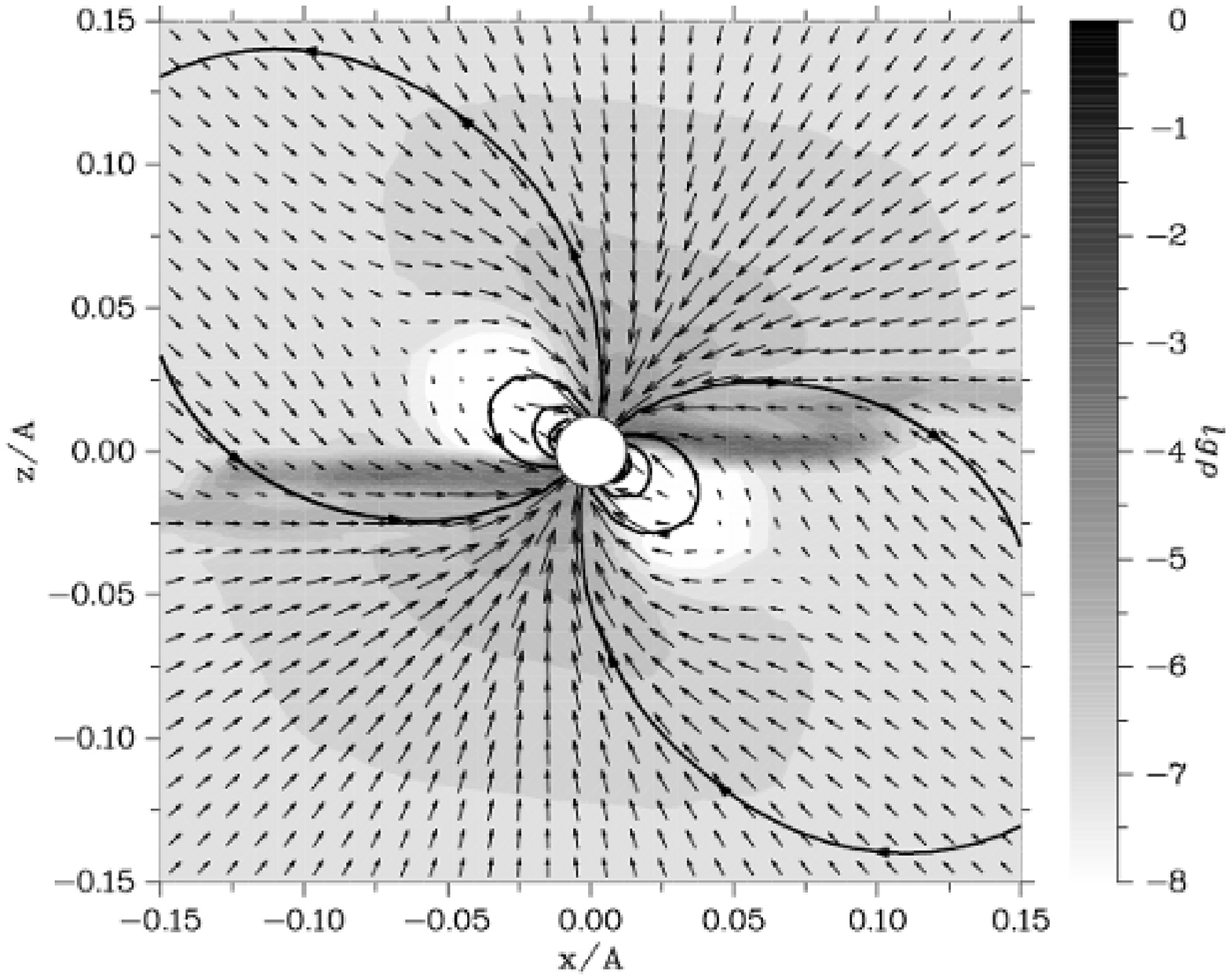}%
\caption{Same as Fig. \ref{fg2} for model 3 ($B_a = 10^6$~G).}%
\label{fg4}
\end{figure}

Figures \ref{fg2}--\ref{fg4} show the flow structures in the equatorial ($xy$) and vertical ($xz$) planes for model 1 ($B_a = 10^5$~G), model 2 ($B_a = {5 \times 10^5}$~G), and model 3 ($B_a = 10^6$~G), respectively. The distributions of the density (contours or gray scale) and velocity (arrows) are shown. The values of the logarithm of the density are indicated on some of the contours (in units of $\rho(L_1)$). The left panels of these figures also show the boundary of the accretor Roche lobe (dashed curves) and a streamline originating from the inner Lagrange point $L_1$. The right panels show the magnetic lines. 

\begin{figure}[ht]
\centering
\includegraphics[width=0.75\textwidth]{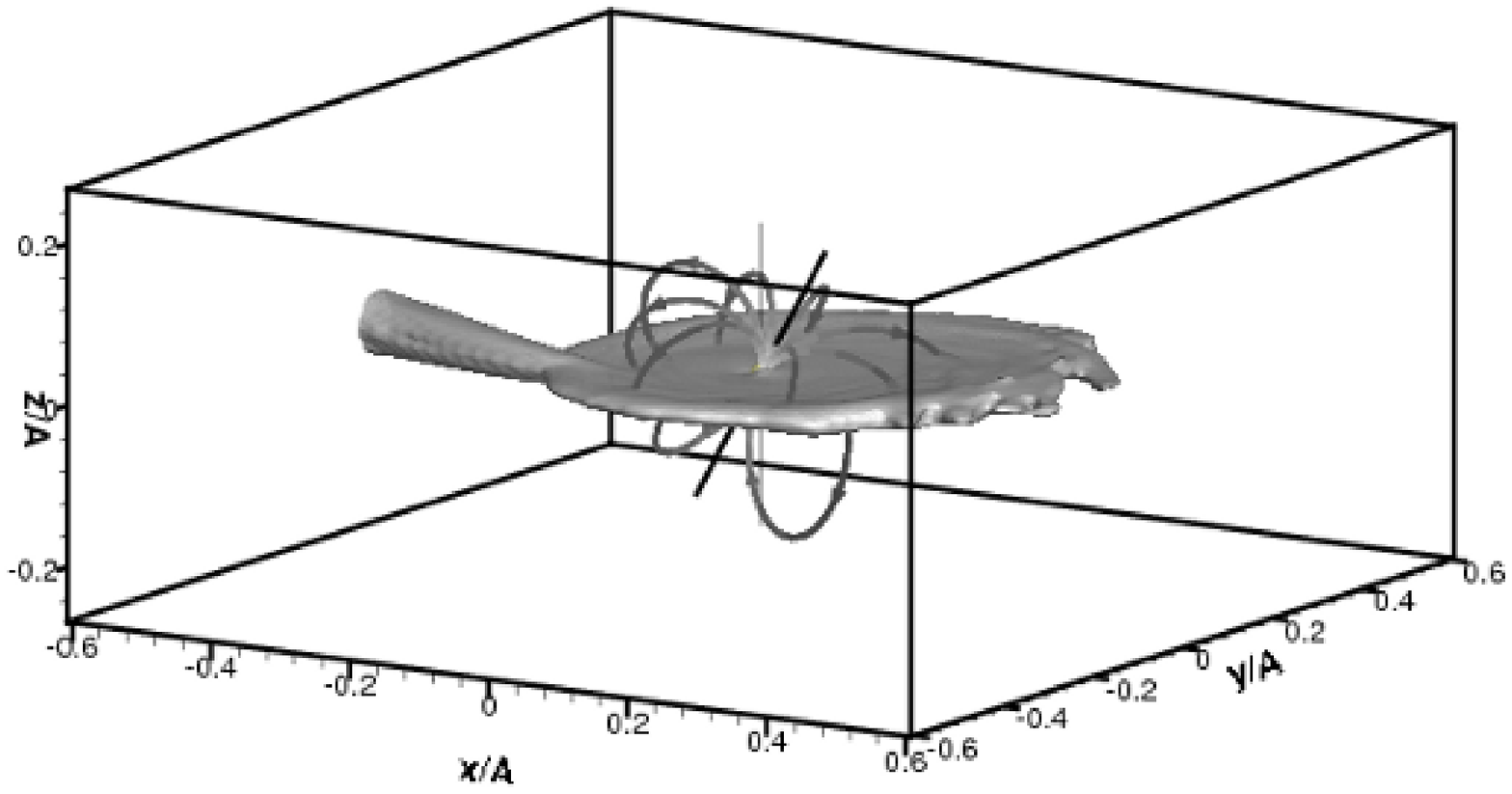}%
\caption{Three-dimensional flow structure for model 1 ($B_a = 10^5$~G). A constant-density surface ($\text{lg}\rho=-4.5$, where $\rho$ is in units of $\rho(L_1)$) and magnetic lines are shown. The gray scale for the magnetic lines displays the magnetic field strength. The thin, vertical line indicates the rotational axis of the accretor, and the bold, inclined line its magnetic axis.}%
\label{fg5}
\end{figure}

\begin{figure}[ht]
\centering
\includegraphics[width=0.75\textwidth]{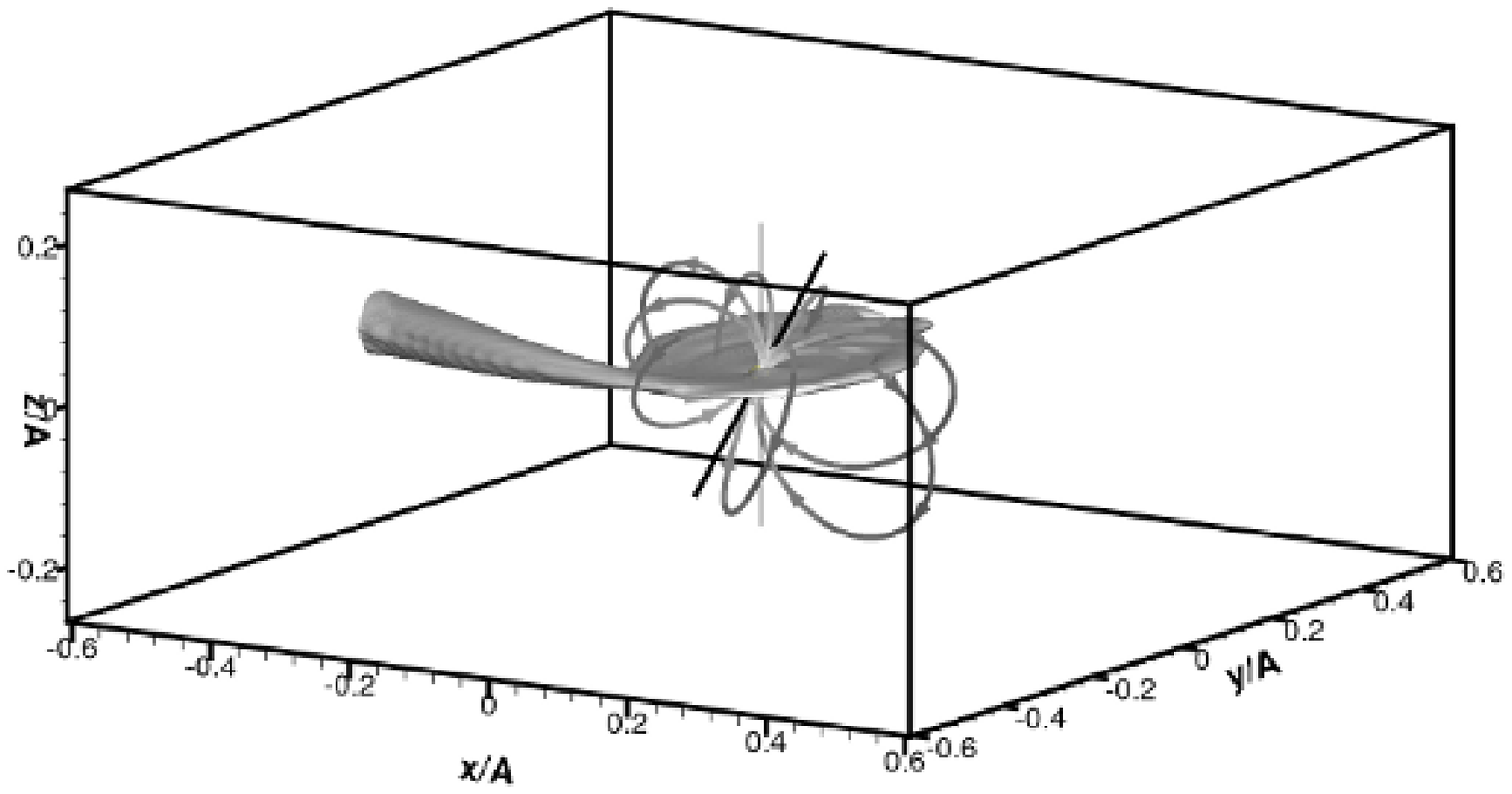}%
\caption{Same as Fig. \ref{fg5} for model 2 ($B_a = {5 \times 10^5}$~G).}%
\label{fg6}
\end{figure}

\begin{figure}[ht]
\centering
\includegraphics[width=0.75\textwidth]{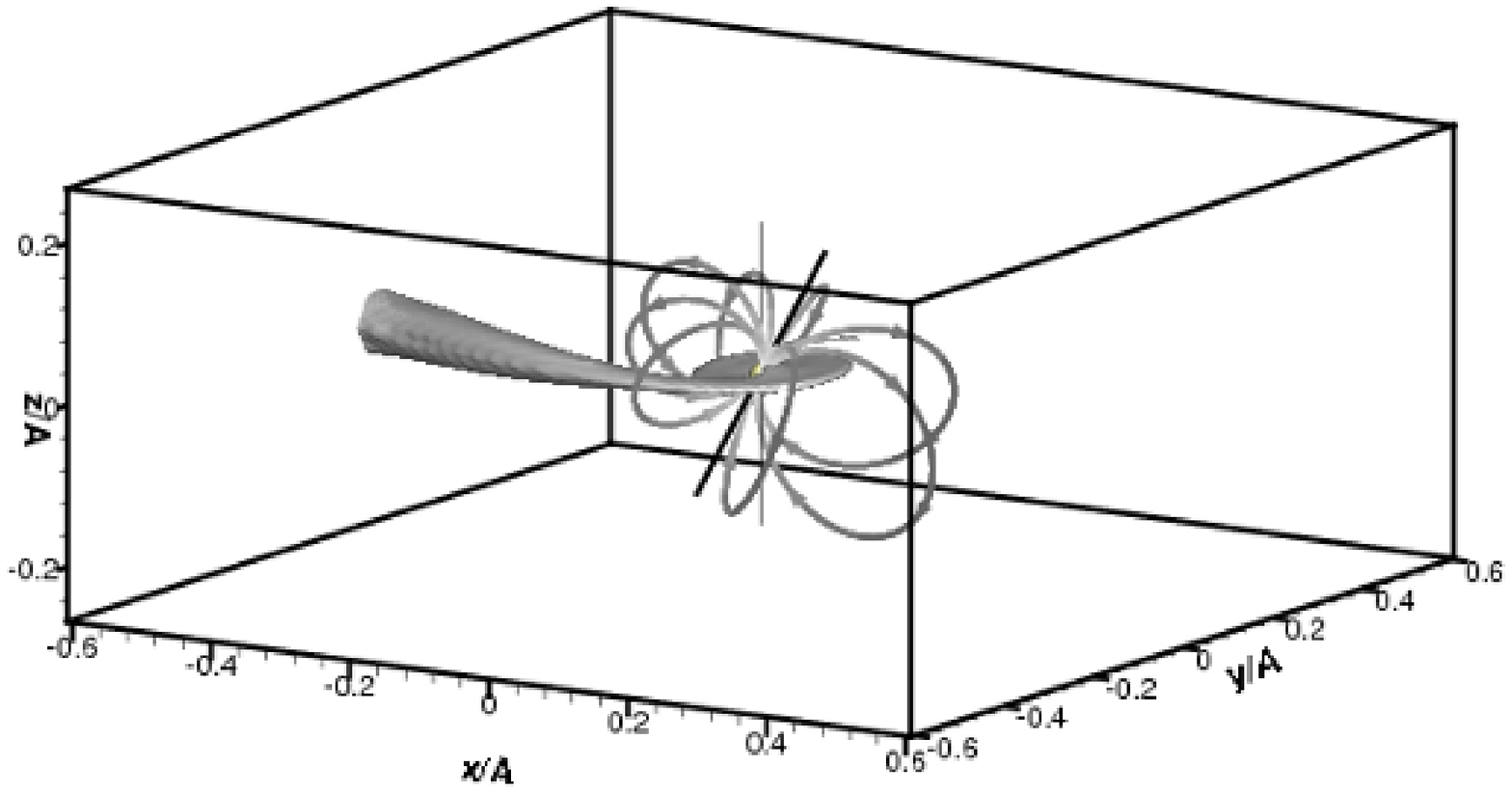}%
\caption{Same as Fig. \ref{fg5} for model 3 ($B_a = 10^6$~G).}%
\label{fg7}
\end{figure}

Figures \ref{fg5}--\ref{fg7} present the three-dimensional structure of the flow for models 1--3. These figures show constant surfaces of the logarithm of the density at the level $\text{lg}\rho=-4.5$ (in units of $\rho(L_1)$)and magnetic lines originating from the accretor surface. The gray scale for the magnetic field displays the magnetic field strength. The rotational axis (thin vertical line) and magnetic axis (bold inclined line) of the accretor are also shown.

The flow pattern obtained for model 1 ($B_a = 10^5$~G) is in good agreement with the results of our earlier computations for this case carried out for a rigorous MHD model \cite{Zhilkin2009}. The characteristic parameters of the accretion disk, and especially its structure, are virtually identical for these two sets of computations. A magnetosphere forms near the accretor surface, and the accretion has a column-like nature. Matter in the magnetosphere moves primarily along magnetic lines and falls onto the star mainly in the region of its magnetic poles, forming accretion columns. Vacuum zones form in the region of the magnetic equator of the accretor. This is due to the fact that the magnetic field prevents matter from penetrating to these regions, since the magnetic lines lie predominantly along the stellar surface near the magnetic equator. We can note some modest differences compared to our previous computations. These are associated with the fact that we developed the model we consider here to simulate plasma flows in strong magnetic fields. A more refined adjustment of parameters is apparently required for the case of relatively weak fields (in particular of $\alpha_w$) in order to obtain more complete agreement with the more rigorous MHD model.

The outer radius of the accretion disk becomes substantially smaller (about $0.15A$) when $B_a = {5 \times 10}^5$~G (model 2). The effective magnetic braking and angular-momentum transfer are enhanced, as is demonstrated by the behavior of the streamline shown in Fig. \ref{fg3}. The magnetosphere region has become much larger. The accretion columns near the magnetic poles of the accretor are more clearly visible.

Finally, the accretion disk essentially becomes degenerate when $B_a = 10^6$~G (model 3). Matter is able to undergo one to two orbital revolutions before falling onto the accretor. The term ''spiral disk'' is more suitable for describing this flow structure, since the matter velocities involved differ strongly from their Keplerian values. The outer radius of this accretion spiral disk is roughly $0.1A$. It is located virtually entirely in the region of the accretor magnetosphere. A substantial part of the spiral disk is comprised of the accretion-column streams. This model essentially corresponds to the limiting case of intermediate polars.

Note that the size of the accretor magnetosphere region obtained in the numerical computations is in good agreement with analytical estimates obtained using \eqref{eq3.2} (Fig. \ref{fg1}).

\subsection{Strong magnetic fields}

\begin{figure}[ht]
\centering
\includegraphics[width=0.5\textwidth]{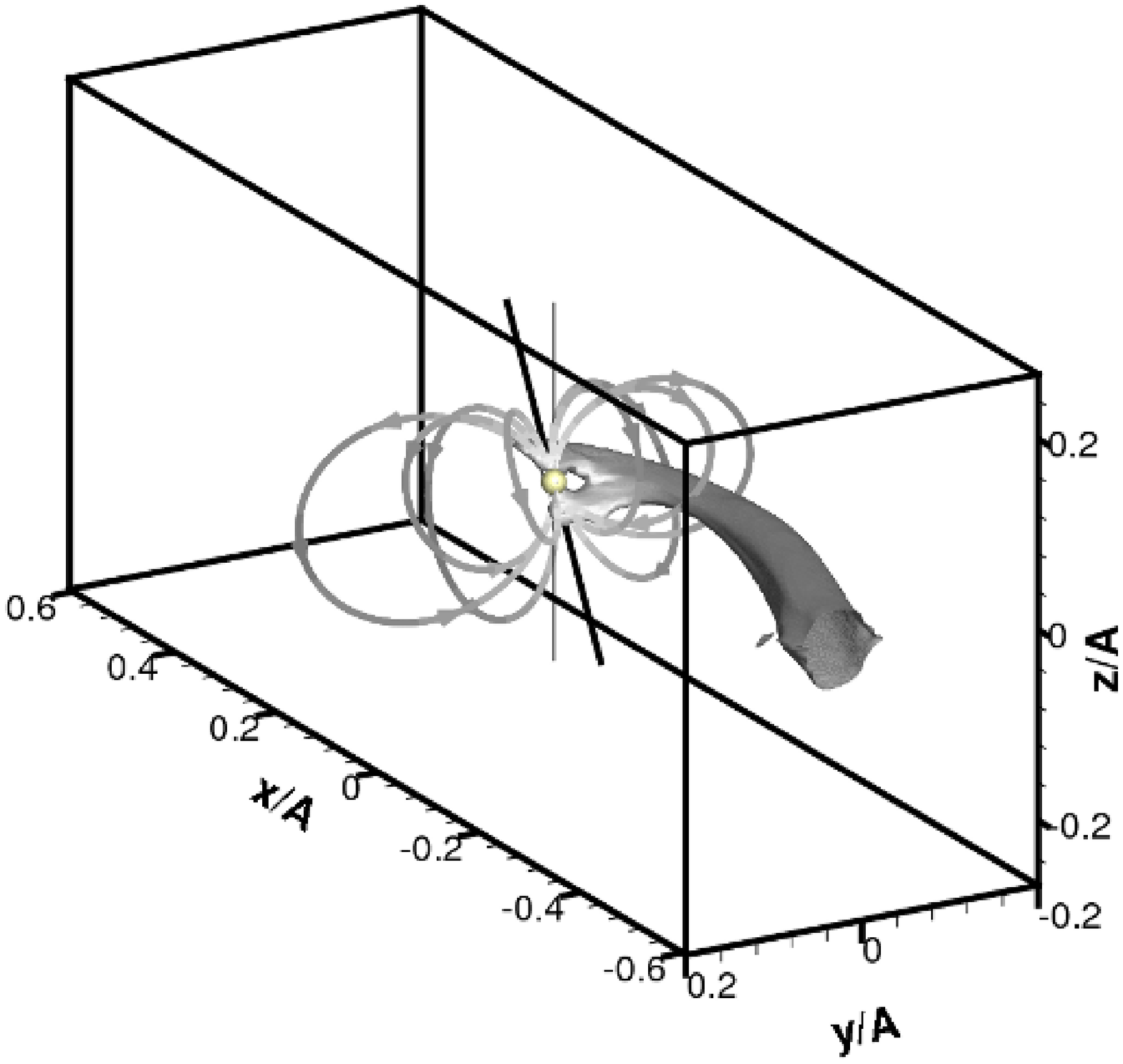}%
\vspace{0.1cm}
\includegraphics[width=0.6\textwidth]{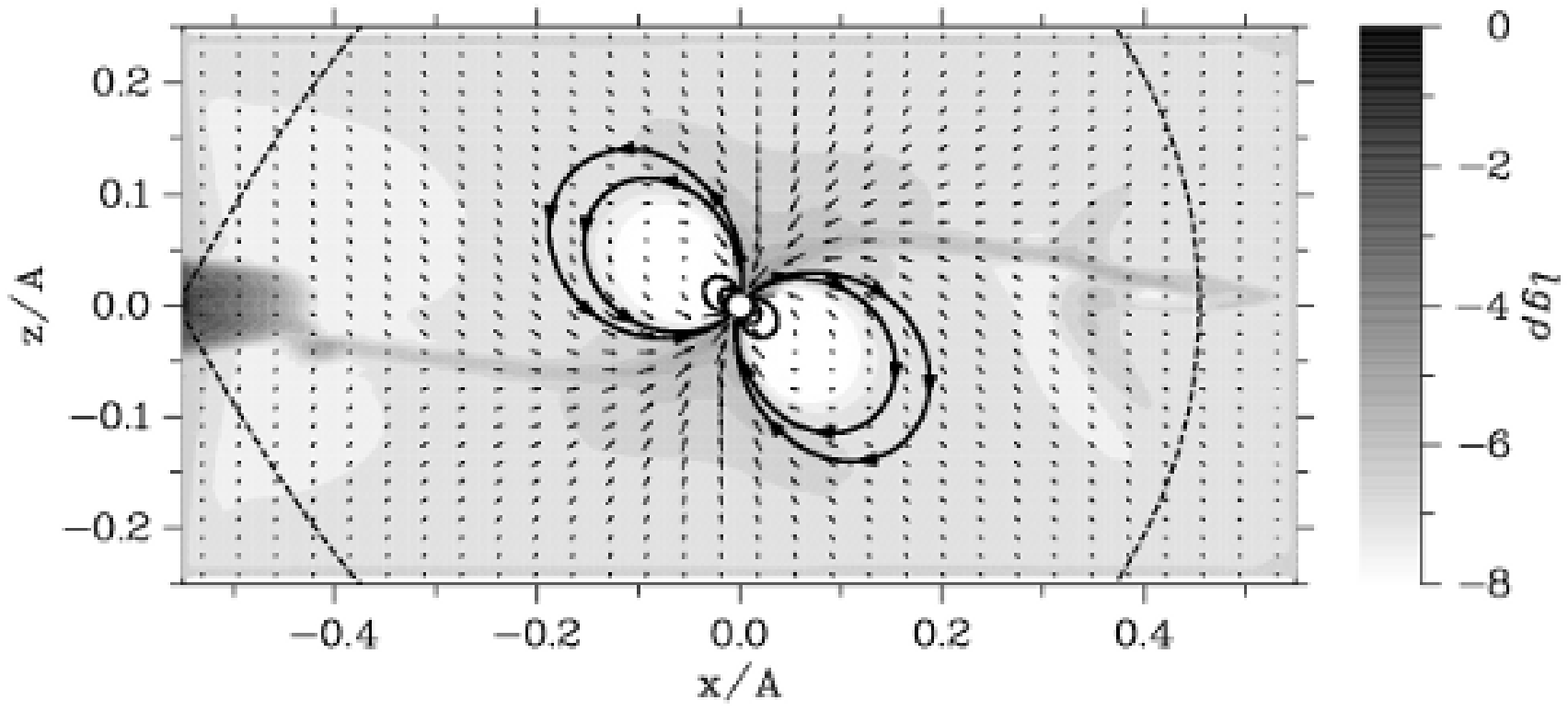}%
\caption{Flow structure in model 4 ($B_a = {5 \times 10}^6$~G). A constant-density surface ($\text{lg}\rho = -5$, where $\rho$ is in units of $\rho(L_1)$) is shown in the upper panel. The remaining notation is the same as in Fig. \ref{fg5}. The lower panel shows the distributions of the density (gray scale) and velocity (arrows) in the vertical plane. The dashed line corresponds to the boundary of the accretor Roche lobe. The bold lines with arrows show magnetic lines.}%
\label{fg8}
\end{figure}

\begin{figure}[ht]
\centering
\includegraphics[width=0.5\textwidth]{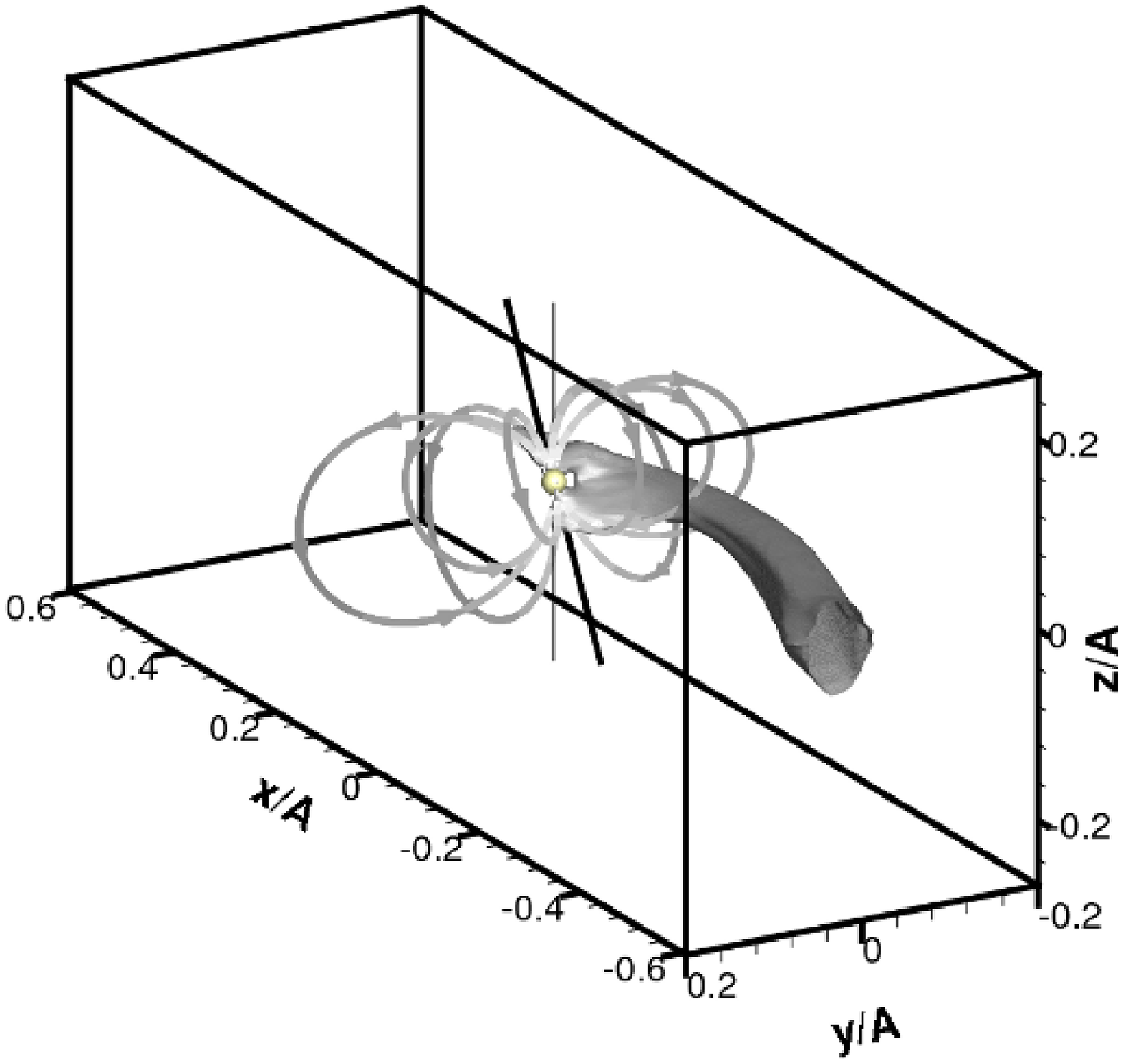}%
\vspace{0.1cm}
\includegraphics[width=0.6\textwidth]{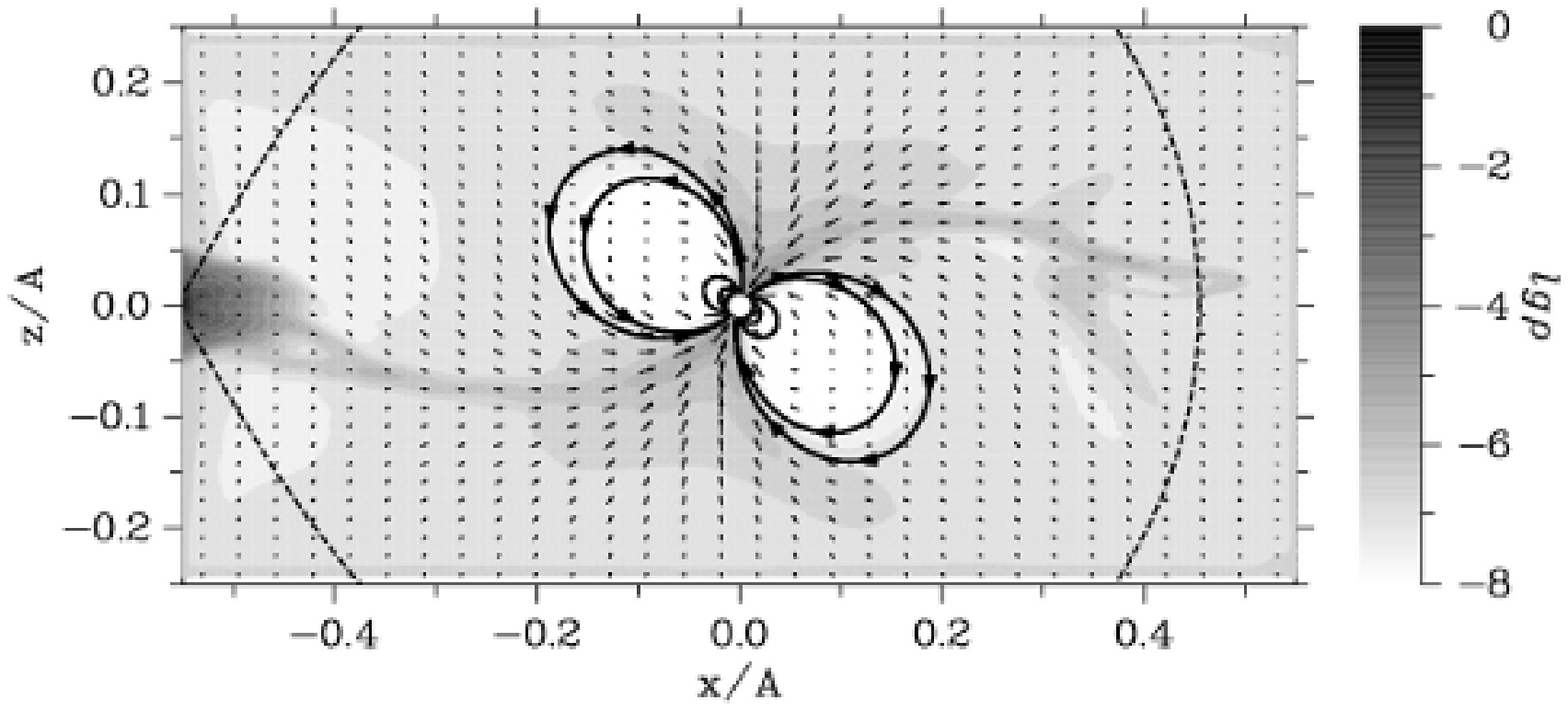}%
\caption{Same as Fig. \ref{fg8} for model 5 ($B_a = 10^7$~G).}%
\label{fg9}
\end{figure}

\begin{figure}[ht]
\centering
\includegraphics[width=0.5\textwidth]{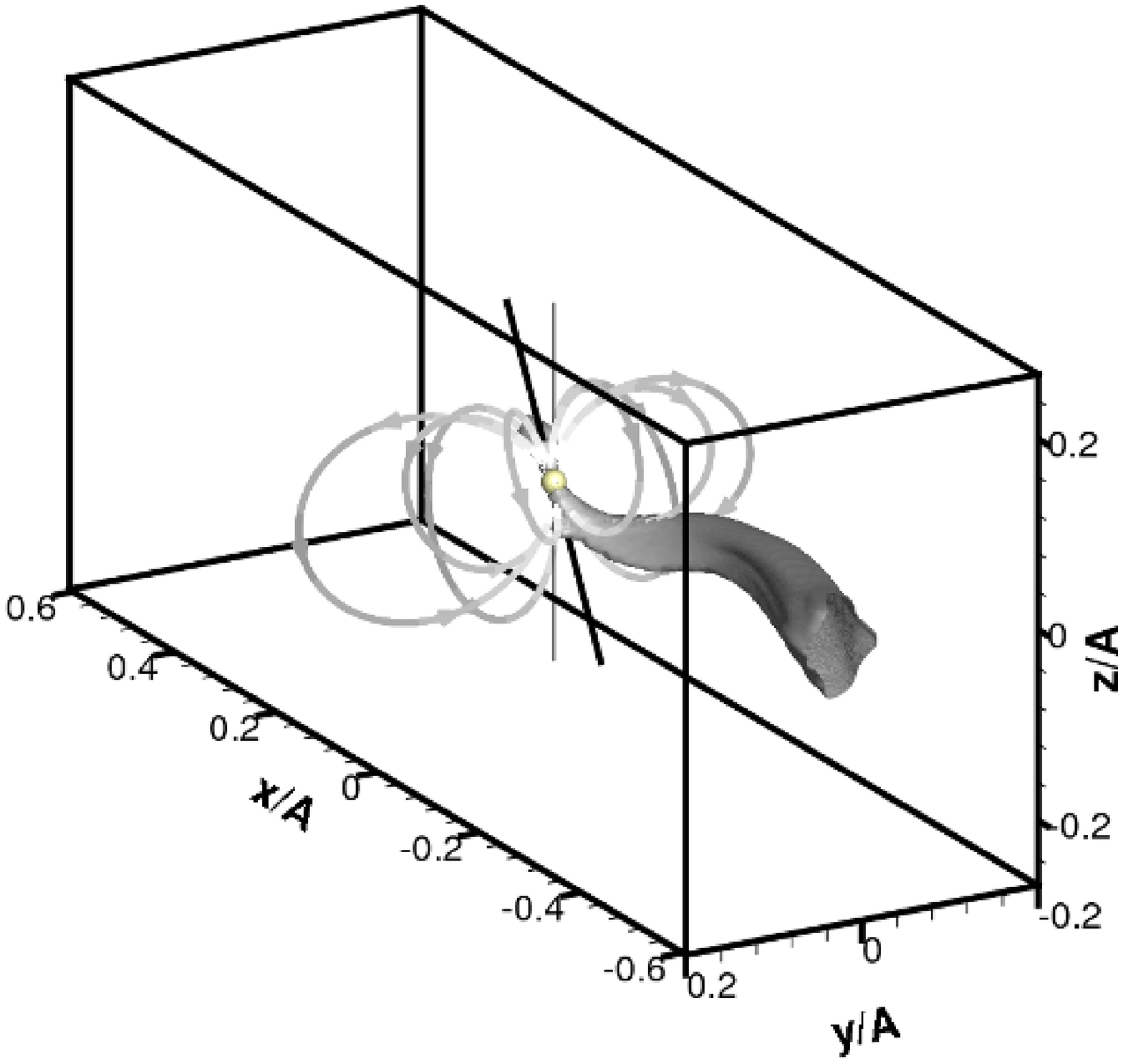}%
\hspace{0.1cm}
\includegraphics[width=0.6\textwidth]{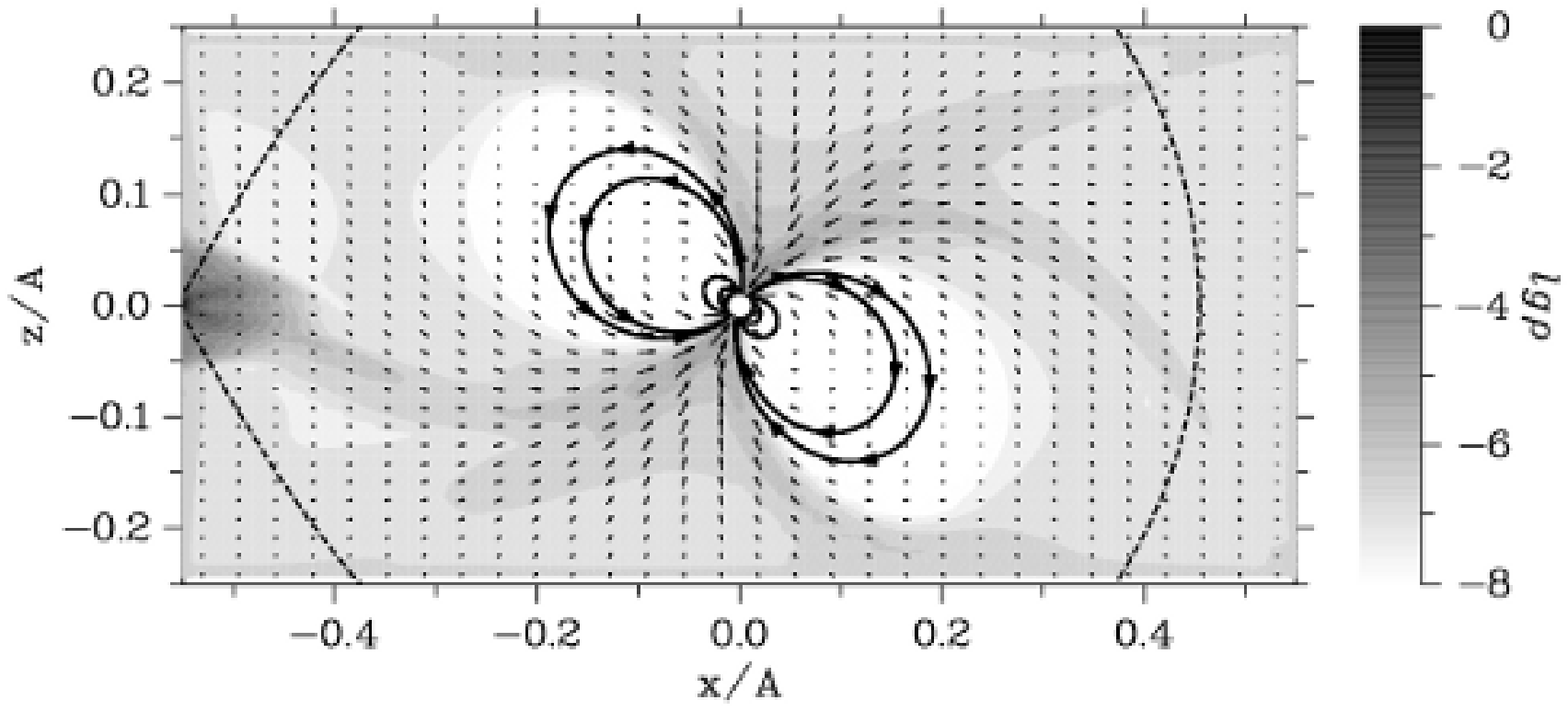}%
\caption{Same as Fig. \ref{fg8} for model 6 ($B_a = {5 \times 10}^7$~G).}%
\label{fg10}
\end{figure}

\begin{figure}[ht]
\centering
\includegraphics[width=0.5\textwidth]{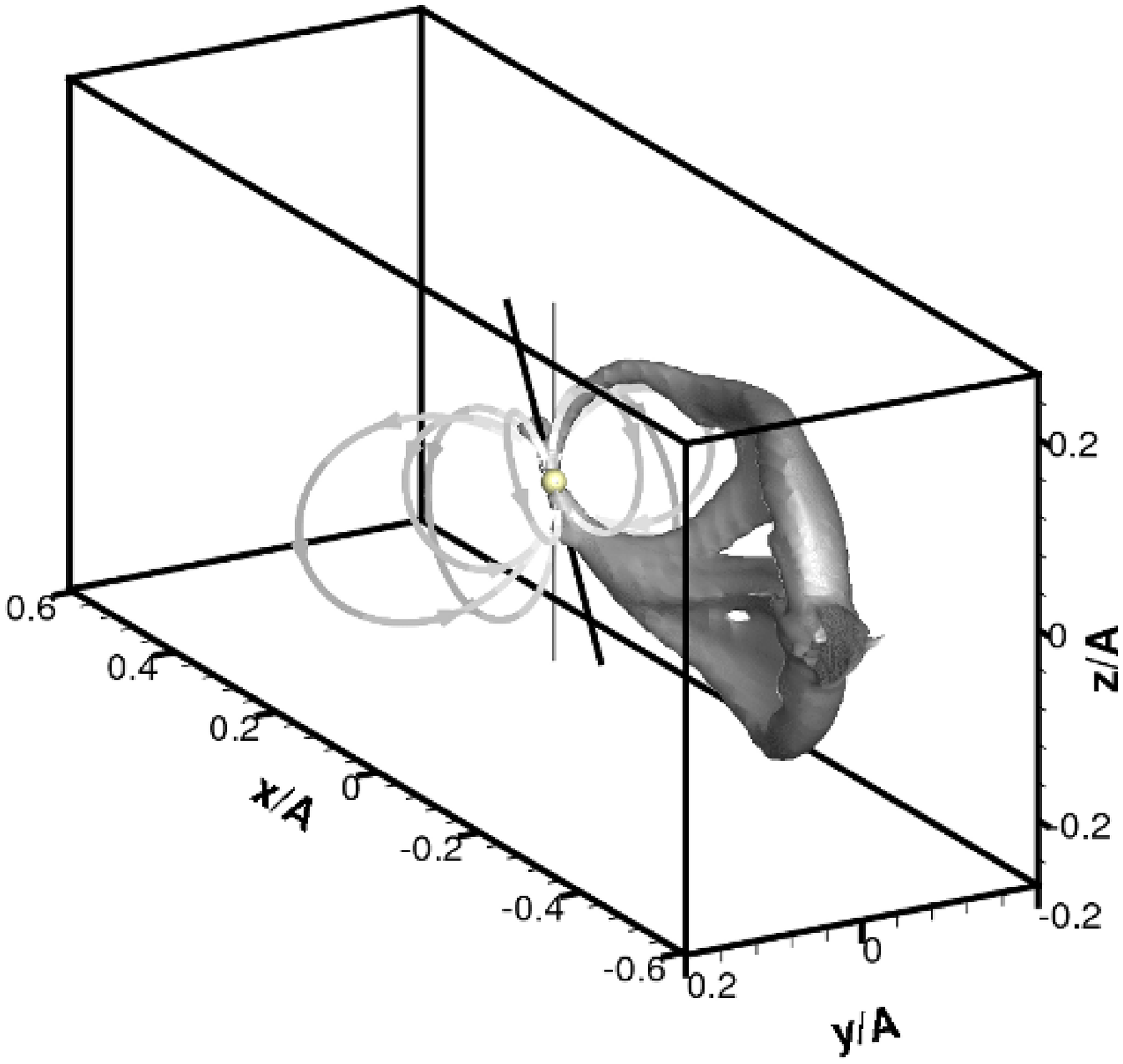}%
\hspace{0.1cm}
\includegraphics[width=0.6\textwidth]{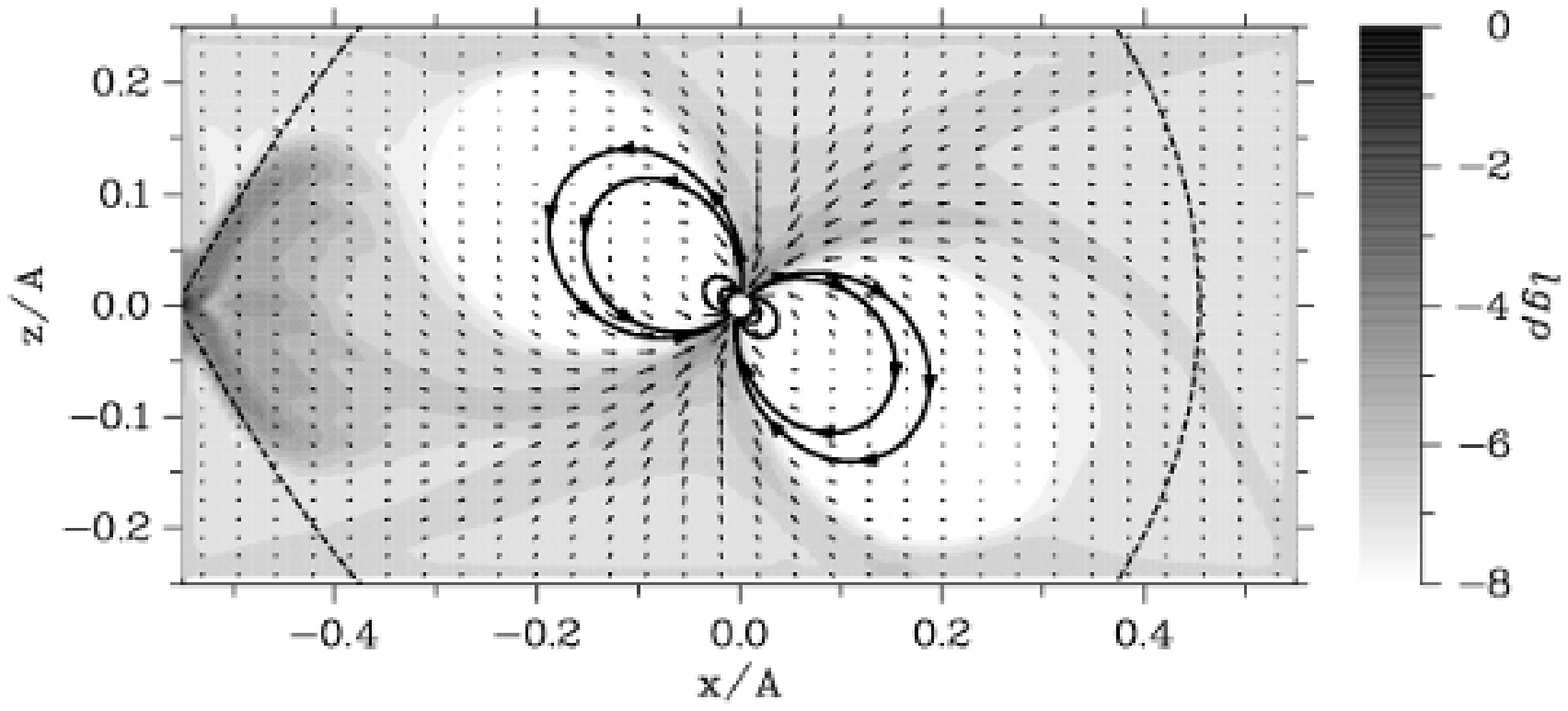}%
\caption{Same as Fig. \ref{fg8} for model 7 ($B_a = 10^8$~G).}%
\label{fg11}
\end{figure}

The MHD flow structures for models 4 ($B_a = {5 \times 10}^6$~G), 5 ($B_a = 10^7$~G), 6 ($B_a = {5 \times 10}^7$~G), and 7 ($B_a = 10^8$~G) are presented in Figs. \ref{fg8}, \ref{fg9}, \ref{fg10} and \ref{fg11}, respectively. 

The upper panels in these figures present the three-dimensional structure of the flow. These show constant surfaces of the logarithm of the density at the level $\text{lg}\rho = -5$ (in units of $\rho(L_1)$) and magnetic lines. The gray scale for the magnetic field displays the magnetic field strength. The rotational axis (thin, vertical line) and magnetic axis (bold, inclined line) of the accretor are also shown. The images have been turned so that features of the accretion flow near the accretor surface are visible. The right panels show distributions of the density (gray scale) and velocity (arrows) in the vertical plane $xz$. The boundary of the accretor Roche lobe (dashed curve) and magnetic lines are also shown.

Analysis of these figures indicates that the flow structures for this group of models are qualitatively different from those for the previous cases (models 1, 2, and 3). No accretion disk forms in these models, and the flow takes the form of accretion-column streams. In models 4 and 5, the streams originating from the inner Lagrange point $L_1$ divide into two flows when they reach the accretor surface, with the left and right flows falling onto the North and South magnetic poles of the star. The northern accretion flow is more powerful in model 4, and the southern in model 5. This comes about because, in our formulation of the problem, the South magnetic pole is closer to $L_1$ than the North magnetic pole. Therefore, the motion of plasma toward the South magnetic pole is energetically more favorable in a stronger magnetic field. This is supported by our results for model 6, in which only one accretion flow forms, ending at the South magnetic pole of the star.

The magnetic field in model 7 is so strong that it virtually completely controls the flow inside the accretor Roche lobe. The matter is captured by the magnetic field essentially immediately behind the Lagrange point, and is directed along the magnetic lines toward the stellar surface, forming a powerful southern stream and weaker northern stream. Figure \ref{fg1} shows that, in this case, the magnetosphere becomes larger than the accretor Roche lobe, and therefore partially encompasses the donor envelope. This model probably corresponds to so-called \emph{magnetors} \cite{Lipunov1987}, which represent the limiting case of a polar whose white dwarf has a very strong magnetic field.

\subsection{Discussion}

Thus, in our computations of the flow structures in close-binary systems whose parameters correspond to SS Cyg, models with magnetic fields $B_a \le 10^6$~G form an accretion disk, while no accretion disk forms in models with stronger fields. We can present the following simple arguments to explain this result. In a first approximation, the behavior of the stream of matter flowing from the inner Lagrange point $L_1$ inside the accretor Roche lobe can be analyzed in a ballistic approximation, without including effects associated with the pressure or magnetic field \cite{Boyarchuk2002, Bisikalo2008}. This approximation is justified by the supersonic nature of the flow in the stream. A trajectory analysis indicates that the stream approaches fairly closely to the accretor surface \cite{Lubow1975}. The minimum distance between the trajectory of an individual particle and the center of the star $R_{\text{min}}$ depends on the mass ratio $q = \sfr{M_d}{M_a}$. In the range $0.05 \le q \le 1$, the quantity $R_{\text{min}}$ can be approximated by the following expression, with accuracy to within 1\% \cite{Warner1995}:
\begin{equation}\label{eq3.3}
 R_{\text{min}} = 0.0488~q^{-0.464} A.
\end{equation}

If $R_{\text{min}}$ is larger than the magnetosphere radius $r_m$ [see \eqref{eq3.2}], the magnetic field does not exert an important influence on the motion of matter. The flow can circle around the star and eventually intersect itself at some point. The further evolution of this flow leads to the formation of an accretion disk in the system. If the minimum distance $R_{\text{min}}$ is smaller than the magnetosphere radius $r_m$, the flow ends up in the zone where the magnetic field exerts a substantial influence at some point in its trajectory. The action of the electro-magnetic forces in this zone leads to deceleration of the flow and a loss of its angular momentum. As a result, the flow is not able to circle around the star and form an accretion disk. Thus, the boundary between intermediate polars (which form accretion disks) and polars (which do not) is determined by the condition $r_m = R_{\text{min}}$. Substituting the parameters for the SS Cyg system into this relation, we find that the magnetic field separating these two regimes should be $B_a \approx 10^6$~G. However, this estimate of the magnetic field separating intermediate polars and polars has a fairly general character, since the values of $q$ vary only slightly in cataclysmic variables, and are roughly equal to $0.5$. 

\begin{figure}[ht]
\centering
\includegraphics[width=0.75\textwidth]{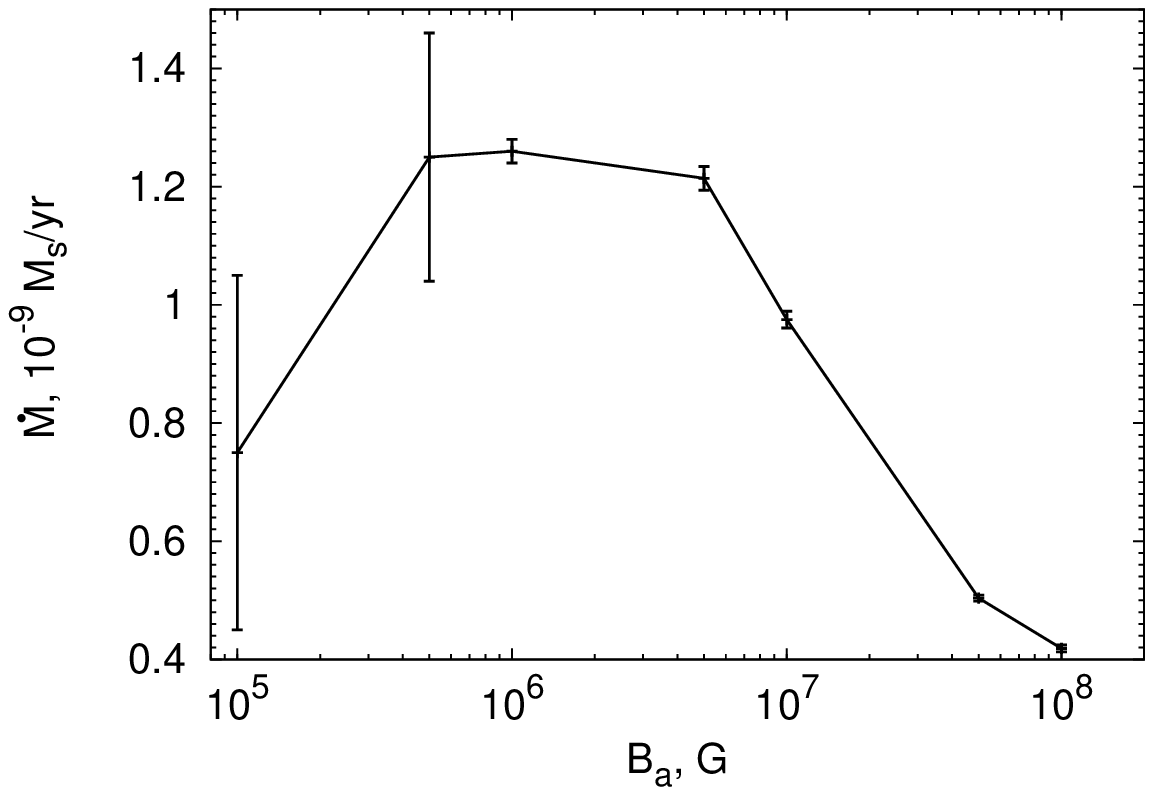}%
\caption{Dependence of the accretion rate on the magnetic field at the accretor surface $B_a$. The vertical line segments show the characteristic amplitudes of the variations of the accretion rate.}%
\label{fg12}
\end{figure}

Figure \ref{fg12} shows the dependence of the accretion rate $\dot{M}_a$ on the magnetic field $B_a$ at the accretor surface obtained in our computations. The vertical line segments illustrate the character of the variations in the accretion rate. The main property of this dependence is its non-monotonic character. The accretion rate grows for fields $B_a < 10^6$~G, and the amplitude of the accretion-rate variations decrease. A maximum accretion rate is reached at $B_a = 10^6$~G. With further increase in $B_a$, the accretion rate falls off. 

This dependence fully corresponds to the reasoning presented above. When $B_a < 10^6$~G, an accretion disk forms in the system, and the accretion rate is determined by angular momentum transfer in the disk. Increasing the magnetic field leads to an increase in the efficiency of magnetic braking in the disk. Therefore, the $\dot{M}_a(B_a)$ dependence acquires a positive slope when $B_a < 10^6$~G. Beginning with a magnetic field of $B_a = 10^6$~G, no accretion disk forms in the system, and the flow acquires the character of accretion-column streams. In this case, the accretion rate $\dot{M}_a$ is determined by the penetrating ability of the stream. The stream cross section decreases and the accretion rate falls off with growth in the magnetic field.

\section{Conclusion}

We have presented a numerical model designed to investigate mass transfer in semi-detached binary systems with strong accretor magnetic fields. Our computations assume the accretor magnetic field to be dipolar, with the magnetic axis inclined to the rotational axis. At the base of the model is the assumption that the plasma dynamics are determined by the slow mean flow, which forms a backdrop for MHD waves that propagate with high velocities. The equations describing the slow motion of the matter were obtained by averaging over rapidly propagating pulsations. The strong external magnetic field plays the role of an effective fluid with which the plasma interacts. In the equation of motion, the mean electro-magnetic force exerted by the accretor magnetic field has a form analogous to the friction force between different components in a plasma consisting of several types of particle. Magnetic diffusion due to current dissipation in turbulent vortices and magnetic buoyancy is included in the numerical model. Moreover, averaging the induction equation over rapidly propagating MHD waves leads to additional dissipation of the magnetic field (wave diffusion).

We have used the model we have developed together with a modified, three-dimensional, parallel numerical code to simulate MHD flow structures in close-binary systems whose parameters correspond to those of SS Cyg. We have presented computational results for various magnetic fields $B_a$ at the accretor surface. An accretion disk forms in the system in models with relatively weak magnetic fields $B_a \le 10^6$~G. As $B_a$ increases, the size of the accretion disk decreases and the magnetosphere radius increases. The accretion disk degenerates into a spiral disk when $B_a = 10^6$~G, in which the matter is able to undergo only one to two orbital revolutions before falling onto the star. These types of flow apparently correspond to the conditions in intermediate polars. No accretion disk forms in models with strong magnetic fields $B_a > 10^6$~G. The flow takes the form of accretion-column streams originating from the inner Lagrange point $L_1$ and ending at the magnetic poles of the accretor. The morphological structure of such flows correspond to polars.

The field $B_a = 10^6$~G separating these two flow regimes is determined by the relationship between the minimum distance to which the stream of matter approaches the accretor $R_{\text{min}}$ and the magnetosphere radius $r_m$. An accretion disk forms when $R_{\text{min}} > r_m$. Otherwise, the flow ends up at some point in the zone near the accretor where the magnetic field exerts a substantial influence, preventing the accretion disk from forming. Note that this estimate of the magnetic field separating intermediate polars and polars has a fairly general character, since it depends only weakly on the parameters of the system.

Our computational results show that the dependence of the accretion rate on $B_a$ is non-monotonic. The accretion rate grows for $B_a < 10^6$~G, reaching its maximum value when $B_a = 10^6$~G, with the accretion rate falling off upon further increase in $B_a$. The increase in the accretion rate for $B_a < 10^6$~G is due to magnetic braking in the accretion disk. When $B_a > 10^6$~G, no accretion disk forms in the system, and the flow acquires the character of accretion-column streams. Therefore, the stream cross section decreases with growth in $B_a$, and the accretion rate falls off in this case.\\

\noindent \textbf{Acknowledgments.} This work was supported by the Basic Research Programme of the Presidium of the Russian Academy of Sciences ''The Origin, Structure, and Evolution of Objects in the Universe'', the Russian Foundation for Basic Research (projects 08-02-00371, 09-02-00064), the Federal Targeted Program ''Science and Science-Education Departments of Innovative Russia in 2009-2013'' (Ministry of Science and Education of the Russian Federation grant ''Studies of Non-stationary Processes in Stars and the Interstellar Medium at the Institute of Astronomy of the Russian Academy of Sciences''). The authors thank S.N. Zamozdra for useful discussions.

\section*{Application}

\appendix

\section{Plasma dynamics in a strong magnetic field}

Plasma dynamics in a strong external magnetic field are characterized by the relatively slow mean motion of particles along the magnetic lines, their drift across the field lines, and the propagation of Alfv{\'e}n and magneto-acoustic waves with very high speeds against the backdrop of these slow motions. Over the characteristic dynamical time scale, the MHD waves are able to traverse the flow region (along a column-like stream, for example) many times. Therefore, we can investigate the mean flow pattern, considering the influence of fast pulsations by analogy with wave MHD turbulence. To describe the slow motion of the plasma itself, it is necessary to separate out rapidly propagating fluctuations and apply a well defined procedure for averaging over the ensemble of wave pulsations.

Let us consider the relation
\begin{equation}\label{eqa1}
 \vec{E} + \fr{1}{c}\left( \vec{v} \times \vec{B} \right) = \fr{\vec{j}}{\sigma},
\end{equation}
which expresses Ohm's law for the plasma in a magneto-gas-dynamical approximation \cite{Landau2001}. Here, $\vec{E}$ is the electric field in the plasma, $\vec{B} = \vec{B}_{*} + \vec{b}$ the total magnetic field, $\vec{j}$ the current density, and $\sigma$ the conductivity. We will represent all dynamical quantities as sums of mean values and fluctuations; for example, $\vec{b} = \langle\vec{b}\rangle + \delta\vec{b}$. Averaging \eqref{eqa1}, we find
\begin{equation}\label{eqa2}
 \langle\vec{E}\rangle + 
 \fr{1}{c}\left( \langle\vec{v}\rangle \times \langle\vec{b}\rangle \right) +
 \fr{1}{c}\left( \langle\vec{v}\rangle \times \vec{B}_{*} \right) +
 \fr{1}{c} \left\langle \delta\vec{v} \times \delta\vec{b} \right\rangle = 
 \fr{\langle\vec{j}\rangle}{\sigma}.
\end{equation}
The last term on the left-hand side can be estimated using the following expression, often applied in dynamo theory (see, for example \cite{Parker1982, Ruzmaikin1988}):
\begin{equation}\label{eqa3}
 \left\langle \delta\vec{v} \times \delta\vec{b} \right\rangle = 
 \alpha \langle\vec{b}\rangle - 
 \eta_w \left(\nabla \times \langle\vec{b}\rangle \right),
\end{equation}
where $\alpha$ is the mean helicity of the flow and $\eta_w$ is the diffusion coefficient for the mean magnetic field due to wave MHD turbulence. We neglect the first term (the $\alpha$ effect), since it describes the relatively weak and slow generation of the mean magnetic field in the accretion disk. Substituting the resulting expression into \eqref{eqa2} yields
\begin{equation}\label{eqa4}
 c\langle\vec{E}\rangle + 
 \langle\vec{v}\rangle \times \langle\vec{b}\rangle +
 \langle\vec{v}\rangle \times \vec{B}_{*} -
 \eta_w \left(\nabla \times \langle\vec{b}\rangle \right) = 
 \eta_{\text{OD}} \left(\nabla \times \langle\vec{b}\rangle \right),
\end{equation}
where $\eta_{\text{OD}} = \sfr{c^2}{(4\pi\sigma)}$ is the Ohmic-diffusion coefficient for the magnetic field.

Let us compare the relative contributions of terms in the resulting expression. Since, as a rule, $\eta_w \gg \eta_{\text{OD}}$, the right-hand side can be neglected. Further, we can neglect the second term compared to the third in the case of strong external magnetic fields. Finally, averaging Maxwell's electro-magnetic induction equation, we obtain
\begin{equation}\label{eqa5}
 \nabla \times \langle\vec{E}\rangle = -\fr{1}{c}\pdiff{\langle\vec{b}\rangle}{t}.
\end{equation}
Thus, the first term in \eqref{eqa4} is associated with variations in the mean magnetic field $\langle\vec{b}\rangle$ over the characteristic dynamical time scale. Therefore, this term corresponds to the second term to order of magnitude. Substituting the dominating terms into \eqref{eqa4} leads the expression
\begin{equation}\label{eqa6}
 \langle\vec{v}\rangle \times \vec{B}_{*} =
 \eta_w \left(\nabla \times \langle\vec{b}\rangle \right).
\end{equation}

This relation can be used to calculate the mean electro-magnetic forces in the equation of motion. Neglecting fluctuations of the density, the wave magnetic pressure and the magnetic tension, we find
\begin{equation}\label{eqa7}
 \fr{\langle \vec{B} \times (\nabla \times \vec{B}) \rangle}{4\pi\rho} =
 \fr{\langle \vec{b} \rangle \times (\nabla \times \langle \vec{b} \rangle)}{4\pi\rho} +
 \fr{\vec{B}_{*} \times (\nabla \times \langle \vec{b} \rangle)}{4\pi\rho}.
\end{equation}
The first term on the right-hand side describes the electro-magnetic force due to the intrinsic magnetic field of the plasma, $\langle\vec{b}\rangle$ (with the opposite sign). The second term can be calculated using \eqref{eqa6}. We then have
\begin{equation}\label{eqa8}
 \fr{\vec{B}_{*} \times (\nabla \times \langle \vec{b} \rangle)}{4\pi\rho} = 
 \fr{\vec{B}_{*} \times (\langle\vec{v}\rangle \times \vec{B}_{*})}{4\pi\rho\eta_w} = 
 \fr{\langle\vec{v}\rangle_{\perp}}{\tau},
\end{equation}
where $\langle\vec{v}\rangle_{\perp}$ is the component of the mean plasma velocity perpendicular to the magnetic field $\vec{B}_{*}$, and the characteristic relaxation time is
\begin{equation}\label{eqa9}
 \tau = \fr{4\pi\rho\eta_w}{B_{*}^2}.
\end{equation}

\section{Some estimates}

The diffusion coefficient for the magnetic field $\eta_w$ associated with wave MHD turbulence can be estimated using the expression
\begin{equation}\label{eqb1}
 \eta_w = \fr{\tau_w}{3} \langle \delta v^2 \rangle,
\end{equation}
where $\tau_w$ is the correlation time for pulsations. This expression can be parametrized as follows:
\begin{equation}\label{eqb2}
 \eta_w = \alpha_w \fr{B_{*} l_w}{\sqrt{4\pi\rho}},
\end{equation}
where $\alpha_w$ is a dimensionless parameter that is close to unity and $l_w$ is the characteristic spatial scale for the pulsations, for which we can use the scale for inhomogeneity in the external magnetic field, $l_w=\sfr{B_*}{|\nabla B_{*}|}$. Since the accretor magnetic field is dipolar in our model, $l_w \approx \sfr{r}{3}$.

Let us now estimate the dependences of $\eta_w$ and $\tau$ on the radius $r$ in a column accretion flow. Assuming a uniform density distribution in the flow, we obtain from \eqref{eqa9}, \eqref{eqb2} $\eta_w \propto r^{-2}$, $\tau \propto r^4$. However, the perpendicular cross section of the flow should change with approach toward the accretor. We can take this effect into account using the conservation of mass, $\rho v S = \text{const}$. The area of the perpendicular cross section $S$ of the flow can be estimated by applying the conservation of magnetic flux, $B_{*} S = \text{const}$. We obtain for a constant flow speed $v$ $\eta_w \propto r^{-3/2}$, $\tau \propto r^{5/2}$. If we estimate the flow speed as being roughly equal to the free-fall velocity, $v \propto r^{-1/2}$, then $\eta_w \propto r^{-1}$, $\tau \propto r^{3}$. Thus, in all cases, the diffusion coefficient $\eta_w$ increases with approach toward the accretor, while the relaxation time $\tau$ decreases. This means that wave MHD turbulence will lead to the virtually complete dissipation of the mean magnetic field of the plasma $\langle\vec{b}\rangle$ near the accretor, and that the motion of matter will occur primarily along the magnetic lines.

\end{document}